\documentclass[useAMS,usenatbib,twocolumn]{mnras}

\usepackage{amsmath}
\usepackage{graphicx}
\usepackage{dcolumn}
\usepackage{multicol}
\usepackage{bm}
\usepackage{epsfig}
\usepackage{amssymb,latexsym,mathrsfs}
\usepackage{graphicx}
\usepackage{enumitem}
\usepackage{color}
\usepackage{hyperref}
\usepackage{bm}
\usepackage{booktabs}

\usepackage{float}
\usepackage{natbib}
\usepackage[utf8]{inputenc}
\usepackage[T1]{fontenc}

\makeatletter
\def\hlinewd#1{%
\noalign{\ifnum0=`}\fi\hrule \@height #1 \futurelet
\reserved@a\@xhline}
\makeatother

\usepackage[dvipsnames]{xcolor}

\hypersetup{
    colorlinks=true,
    linkcolor=red,
    citecolor=blue,
}

\def\apjl{{ ApJL}}
\def\apjs{{ ApJS}}

\def\aap{{ A\&A}}

\newcommand{\be}{\begin{equation}}
\newcommand{\ee}{\end{equation}}
\newcommand{\bs}{\begin{split}} 
\newcommand{\bea}{\begin{eqnarray}}
\newcommand{\eea}{\end{eqnarray}}

\title[Magnetar wind nebula explains FRB radio emission]{Quasi-steady emission from repeating fast radio bursts can be explained by magnetar wind nebulae}

\author[Bhattacharya et al.]{
Mukul Bhattacharya$^{1,2,3}$\thanks{mukul.b@utexas.edu}, Kohta Murase$^{3,4}$, Kazumi Kashiyama$^{5}$\\ 
${}^1$Department of Physics, Wisconsin IceCube Particle Astrophysics Center, University of Wisconsin, Madison, WI 53703, USA\\
${}^2$Department of Astronomy, Astrophysics and Space Engineering, Indian Institute of Technology Indore, Simrol, MP 453552, India\\
${}^3$Department of Physics; Department of Astronomy \& Astrophysics; Center for Multimessenger Astrophysics, Institute for Gravitation\\ and the Cosmos, The Pennsylvania State University, University Park, PA 16802, USA\\
${}^4$Center for Gravitational Physics and Quantum Information, Yukawa Institute for Theoretical Physics, Kyoto University, Kyoto\\ 606-8502, Japan\\
${}^5$Astronomical Institute, Graduate School of Science, Tohoku University, Sendai 980-8578, Japan
} 

\begin{document}

\date{Accepted . Received ; in original form }

\pagerange{\pageref{firstpage}--\pageref{lastpage}} \pubyear{2025}

\maketitle

\label{firstpage}

\begin{abstract} 
Among more than 1000 known fast radio bursts (FRBs), only five sources -- FRBs 20121102A, 20190520B, 20201124A, 20240114A and 20190417A -- have confirmed associations with persistent radio sources (PRS). The observed quasi-steady emission is consistent with synchrotron radiation from a composite of magnetar wind nebula (MWN) and supernova (SN) ejecta. 
Using a phenomenological model that incorporates simplified treatments of the nebular dynamics and particle acceleration, we compute the synchrotron flux by solving kinetic equations for energized electrons, accounting for electromagnetic cascades of electron-positron pairs interacting with nebular photons. Within the framework of our model, the rotation-powered scenario requires a young neutron star (NS) with age $t_{\rm age}\approx 20\,{\rm yr}$, dipolar magnetic field $B_{\rm dip}\approx (3-5)\times10^{12}\,{\rm G}$ and initial spin period $P_i\approx 1.5-3\,{\rm ms}$ in an ultra-stripped SN progenitor to account for emissions from FRBs 20121102A and 20190520B. In contrast, FRB 20201124A requires $t_{\rm age}\approx 10\,{\rm yr}$, $B_{\rm dip}\approx 5.5\times10^{13}\,{\rm G}$ and $P_i\approx 10\,{\rm ms}$ in a conventional core-collapse SN progenitor. For the magnetar-flare-powered model, NS aged $t_{\rm age} \approx 25\,/40\,{\rm yr}$ in a USSN progenitor and $t_{\rm age} \approx 12.5\,{\rm yr}$ in a CCSN progenitor explains the observed flux for FRB 20121102A/20190520B and FRB 20201124A, respectively. 
Finally, we estimate a minimum NS age $t_{\rm age,min} \sim 1-3\,{\rm yr}$ based on the near-source plasma contribution to observed DM, and $t_{\rm age,min} \sim 6.5-10\,{\rm yr}$ from the absence of radio signal attenuation.  
\end{abstract}


\section{Introduction}
Fast radio bursts (FRBs) are energetic millisecond duration pulses of coherent emission, located at cosmological distances, whose physical origin is still debated after a decade since their discovery \citep{Lorimer2007,Thornton2013,Spitler2014,CC2019}. Many theoretical models have been proposed to explain the nature of their progenitors, both for repeating and one-off events (see \citealt{Platts2019}, for a recent review). The detection of Galactic FRB 200428 and its association with SGR J1935+2154 \citep{Bochenek2020,CHIME2020}, suggests that FRBs can originate from magnetars born from the core collapse of massive stars (see e.g., \citealt{Murase2016,Kumar2017,MB2019,Kumar2020,Lu2022}).

In addition to FRBs, rapidly rotating pulsars and magnetars have been proposed as central engines of super-luminous supernovae (SNe), stripped-envelope SNe and rapidly-rising optical transients~\citep{Metzger2015,Kashiyama2016,Hotokezaka2017,MM2018}. 
Pulsar wind nebulae (PWNe) are known to be efficient particle accelerators and broadband non-thermal emission has been observed from the nebulae of Galactic PWNe~\citep{TT2010,TT2013}. 
\citet{Murase2016} proposed quasi-steady synchrotron emission as counterparts of both FRBs and pulsar/magnetar-driven SNe (including SLSNe), suggesting efficient conversion of rotation and/or magnetic energy to particle energies within the nascent MWN of a young NS~(e.g.,~\citealt{GSlane2006}). 

Precise localisation of FRBs can provide meaningful insights on their sources, by identifying plausible multi-wavelength counterpart(s) and revealing information about the central engine and its surrounding environment~\citep{Michilli2018}. However, until date, only five confirmed FRB-PRS associations have been made in the sample of localized FRBs; namely those of FRBs 20121102A~\citep{Chatterjee2017}, 20190520B~\citep{Niu2022}, 20201124A~\citep{Bruni2023}, 20240114A~\citep{Bruni2024} and 20190417A~\citep{Moroianu2025}. 
\citet{Yang2016} proposed that persistent radio emission from such sources can be produced due to synchrotron-heating process in a self-absorbed magnetar-powered nebula.

PRSs are continuum radio sources that are significantly bright ($L_{\rm PRS}>10^{29}\,{\rm erg\,s^{-1}\,Hz^{-1}}$) and compact ($R_{\rm PRS} < 1\,{\rm pc}$), for them to be related to ongoing star formation in their host galaxy~\citep{Nimmo2022,Dong2024}. The polarisation level is found to be significantly different between the FRB 20121102A bursts and its PRS emission~\citep{Michilli2018,Plavin2022}, thereby ruling out the possibility that these emissions are of the same intrinsic nature. Furthermore, there is no clear evidence for either repeaters or non-repeaters to be preferentially associated with PRSs~\citep{Law2022}, indicating that these are two separate aspects of the central engine and its environment~\citep{Bhandari2023}. 

Apart from their actively repeating behaviour and association with a compact PRS, these FRB sources also exhibit large values of host galaxy dispersion measure (${\rm DM} \sim 150-1200\,{\rm pc\,cm^{-3}}$) and rotation measure (${\rm RM} \sim 10^3-10^5\,{\rm rad\,m^{-2}}$), indicating the presence of a dense magneto-ionic environment near the PRS emission region. 
If the observed RM primarily arises from the PRS region, the PRS luminosity should be correlated with the large source RM~\citep{Yang2020}. 
Recently,~\citet{Bruni2023,Bruni2024} confirmed such a correlation with the detection of less luminous PRSs associated with FRBs 20201124A and 20240114A, both having ${\rm RM}\sim10^2-10^3\,{\rm rad\,m^{-2}}$. They suggest that for lower values of RM, the PRS radio luminosity will likely fall below the detection threshold of current radio telescopes. 

Energy injection into the MWN surrounding the central NS can occur due to either the rotational energy of a young NS that spins down over time~\citep{CW2016,Connor2016,Lyutikov2016,KM2017}, or the release of magnetic energy from NS interior due to flares originating close to the magnetar~\citep{Lyubarsky2014,Kumar2017,Beloborodov2017,MM2018,ZW2021}. The injected energy continually drives the expansion of MWN out to the SN ejecta, and the persistent emission is powered by relativistic electrons heated at the termination shock of the magnetar wind. Efficient conversion of NS rotation/magnetic energy to particle energy in the termination shock region is required to explain the observed quasi-steady radio emission. 
For rapidly rotating young NS, rotational energy is the primary reservoir that powers the wind nebula as is the case with Galactic PWNe~\citep{TT2010}. However, for a decades-old magnetar, the NS magnetic energy may be more significant in comparison to its rotational energy, as proposed by~\citet{Beloborodov2017} to explain the large RM and PRS luminosity of FRB 20121102A. 

The inferred host galaxy DM contributions for FRBs 20190520B and 20190417A are ${\rm DM}_{\rm host} \simeq 900-1200\,{\rm pc\,cm^{-3}}$~\citep{Niu2022}, which are almost five times larger than that of typical FRB host galaxies~\citep{James2022}. \citet{Zhang2020} used IllustrisTNG simulations to show that the DM contribution from FRB 20190520B-like host galaxies at $z \approx 0.2$ is ${\rm DM}_{\rm host} \approx 150\pm100\,{\rm pc\,cm^{-3}}$, which indicates a significant near-source DM contribution, possibly due to an expanding young SN remnant~\citep{YZ2017,PG2018,ZW2021,Katz2022}. Furthermore, FRB 20190520B's DM decreases with time at the rate $-0.09\pm0.02\,{\rm pc\,cm^{-3}}\,{\rm day}^{-1}$~\citep{Niu2022}, in contrast to the nearly fixed DM of FRB 20121102A~\citep{Hessels2019,Oostrum2020,Li2021}. Such large near-source DM together with its decreasing trend can be explained by an expanding shocked shell of SN remnant~\citep{PG2018,Katz2022}, whereas the large and decreasing RM~(e.g., for FRB 20121102A,~\citealt{Hilm2021}) arises due to the radiative cooling of electron-positron pairs injected into the magnetized nebula.

In this work, we propose that FRB sources with an associated PRS are powered by young magnetars embedded in a composite of MWN and SN remnant. Observed quasi-steady emission is generated by synchrotron radiation of the MWN, with the dense SN ejecta contributing a large fraction of the near-source DM. This paper is organised as follows. In Section~\ref{Sec2}, we discuss the physical model for energy injection from the rotation/magnetar-flare-powered wind region into the magnetized nebula. In Section~\ref{Sec3}, we describe the combined evolution of the MWN and SN ejecta, used to compute synchrotron emission from the magnetized nebula. In Section~\ref{Sec4}, we present the physical constraints needed to explain the detectable radio emission from the PRSs. Based on the observed radio spectral energy distribution (SED), light curve and time evolution of the near-source DM for these sources, we constrain the magnetar parameters and that of the SN ejecta in Section~\ref{Sec5}. We also compare the magnetar parameters derived for the rotation- and magnetar-flare-powered models. We discuss our main results with their implications in Section~\ref{Sec6} and conclude with a summary in Section~\ref{Sec7}. 
Throughout this work, we use the notation $Q = 10^x Q_x$ in CGS units, unless noted otherwise.

\section{Physical model}
\label{Sec2} 
\begin{figure}
\centering
\includegraphics[width=\linewidth,height=5.2cm]{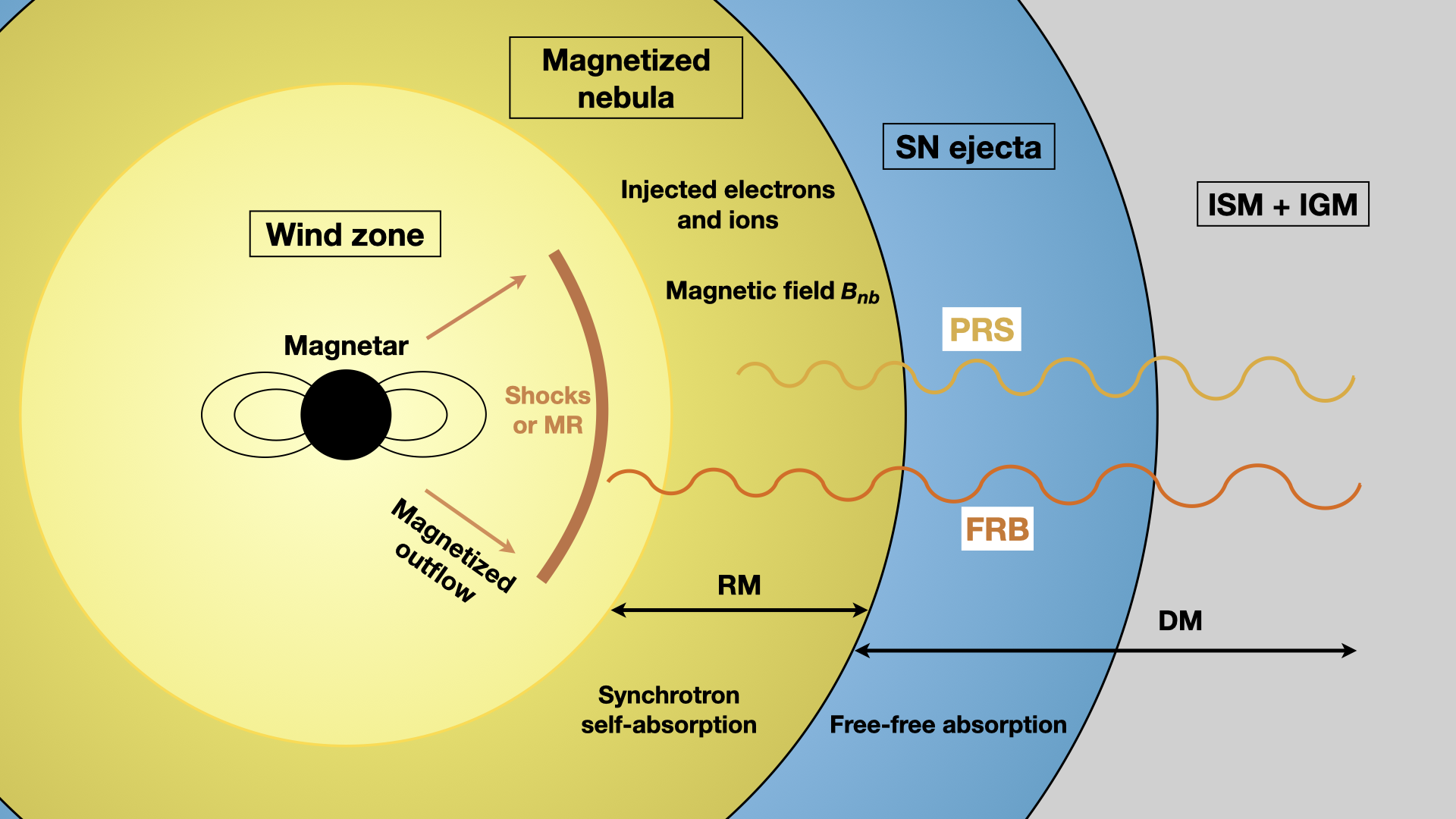}
\vspace{-0.3cm}
\caption{Schematic picture of a magnetar surrounded by magnetized wind nebulae and baryonic SN ejecta. Magnetar flares and/or rotationally powered outflows inject particles and magnetic energy into the nebula. Synchrotron radiation, observed as PRS associated with the FRB, is emitted by energetic electron-positron pairs gyrating within the magnetized nebula. The radio emission is observable once the system becomes optically thin to various processes.
}
\label{fig:frb_schematic}
\end{figure}

A highly magnetised pulsar (or magnetar) is likely to remain as a compact remnant once a SN explosion takes place. The rotational and/or magnetic energy extracted by the outgoing relativistic wind is injected into the associated nebula which then powers the synchrotron emission from the SN ejecta, detectable at radio frequencies.  
Figure~\ref{fig:frb_schematic} shows a schematic diagram of a rapidly rotating young magnetar surrounded by a magnetized nebula and SN ejecta. The spindown/magnetic energy powers outflows that are accelerated in the wind zone between the NS light cylinder and the nebula. The energized electrons and positrons once injected into the nebula gyrate along the magnetic fields to drive the persistent radio emission that is associated with the FRB. The resultant non-thermal emission is powered by synchrotron radiation in the magnetized nebula and SN ejecta. While most of the RM originates from the magnetic fields in the nebula, bulk of the near-source DM is accumulated in the SN ejecta. FRB persistent emission is observable once the system becomes optically thin to synchrotron self-absorption in the nebula and free-free absorption in the ejecta.

\subsection{Energy source of quasi-steady emission}
\citet{Murase2016} considered a young pulsar/magnetar-driven SN as the source of synchrotron emission from wind nebulae, and predicted persistent radio emission as counterparts of FRBs earlier than the discovery of FRB 20121102A. Here we extend their model to explain the persistent radio emission from three sources, namely FRBs 20121102A, 20190520B and 20201124A. We first discuss the energy injection mechanism for these sources, considering both rotation- and magnetic-powered scenarios.

The rotation-powered model has been studied extensively in the context of Crab Nebula (e.g., \citealt{Connor2016,CW2016,Lyutikov2016}). The intrinsic energy arising from NS rotation is $\mathcal{E}_{\rm rot,i} = 0.175 M_{\rm ns} R_{\rm ns}^2 (2\pi/P_i)^2 \approx (1.9\times10^{52}\, {\rm erg})\, P_{i,-3}^{-2}$, where $M_{\rm ns}=1.4\, M_{\odot}$ is the NS mass, $R_{\rm ns}=12\, {\rm km}$ is the NS radius and $P_i$ is its initial spin period. The rotation energy can be extracted by unipolar induction. The initial spindown luminosity is given by
\begin{eqnarray}
\label{L_sd}
    L_{\rm sd,i} \approx \frac{B_{\rm dip}^2 (2\pi/P_i)^4 R_{\rm ns}^6}{4c^3} (1 + {\rm sin}^2 \chi) \nonumber\\ \sim (2.4\times10^{45}\,{\rm erg/s})\,B_{\rm dip,13}^{2} P_{\rm i,-3}^{-4}
\end{eqnarray}
where $B_{\rm dip}$ is the dipolar magnetic field and $\chi$ is the angle between the axes of rotation and dipole magnetic field~\citep{Gruzinov2005,Spitkovsky2006,Tchek2013}. We assume that $B_{\rm dip}$ and ${\rm sin}^2 \chi  = 2/3$ are time independent. The corresponding spindown timescale is $t_{\rm sd} \approx (0.25\, {\rm yr})\, B_{\rm dip,13}^{-2} P_{\rm i,-3}^2$. The spindown luminosity is almost constant for $t \lesssim t_{\rm sd}$, whereas it evolves as $L_{\rm sd} \approx L_{\rm sd,i} \times (t/t_{\rm sd})^{-2}$ for $t \gtrsim t_{\rm sd}$.

Magnetar powered models are promising as evidenced by the recent detection of FRB 200428 from a Galactic magnetar \citep{Bochenek2020,CHIME2020}. The magnetic energy $\mathcal{E}_{\rm B,int} \approx B_{\rm int}^2 R_{\rm ns}^3/6 = (3\times10^{49}\, {\rm erg})\, B_{\rm int,16}^2$ for a given interior magnetic field $B_{\rm int}$, is expected to be more significant in comparison to the rotation energy for a NS with age $t_{\rm age} \gtrsim {\rm few}\,\times 10\, {\rm yr}$. Comparing the minimum energy, $\mathcal{E}_{\rm FRB,min} \approx f_b \mathcal{F}_{\rm FRB} \mathcal{R}_{\rm FRB}t_{\rm age}$, needed to explain a repeating FRB with magnetic energy $\mathcal{E}_{\rm B,int}$ provides a lower limit $B_{\rm int} \gtrsim (10^{15}\, {\rm G})\, f_b^{1/2}\mathcal{F}_{\rm FRB,40}^{1/2}\mathcal{R}_{\rm FRB,-3}^{1/2}t_{\rm age,9.5}^{1/2}$ (see \citealt{KM2017}). Here $f_b$ is the beaming factor, $\mathcal{F}_{\rm FRB}$ is the FRB fluence, $\mathcal{R}_{\rm FRB}$ is the repetition rate for FRB and $t_{\rm age}$ is the NS age. 

Magnetic energy injection has been proposed to explain the PRS associated with FRB 20121102A along with its large rotation measure.~\citet{MM2018} considered a power-law energy injection rate given by 
\begin{equation}\label{Lmag}
    L_{\rm mag} = (\alpha - 1)\frac{\mathcal{E}_{\rm B,int}}{t_{\rm inj}} \left(\frac{t}{t_{\rm inj}}\right)^{-\alpha},
\end{equation}
where $t_{\rm inj} \approx 0.6\, {\rm yr}$ is the time when the onset of energy injection occurs and $\alpha \approx 1.3$ is the power-law index. Here $t_{\rm inj}$ is determined by the timescale for magnetic flux to start leaking out of the magnetar core. \citet{Beloborodov2017} proposed that $t_{\rm inj}$ from flares can be comparable to the pair freeze-out timescale $t_{\pm} \sim 0.1\,{\rm yr}$. Although the energy injection index $\alpha$ can depend on the nature of the FRB source,~\citet{MM2018} showed that $\alpha \gtrsim 1$ adequately explains the radio SED of the PRS associated with FRB 20121102A. This implies that the rate of FRB activity slows down over time or the flares become less energetic on average. Although $L_{\rm mag}$ here is modeled with a smooth function of time, the energy release from magnetars can also occur intermittently. For time $t < t_{\rm inj}$, energy injection is dominated by rotation and the total injected energy is $\mathcal{E}_{\rm tot}(t<t_{\rm inj}) = \int_0^t L_{\rm sd}(t)dt$. In contrast, the interior magnetic energy leaks out into the nebula for $t \geq t_{\rm inj}$, and therefore $\mathcal{E}_{\rm tot}(t \geq t_{\rm inj}) = \int_0^t [L_{\rm sd}(t) + L_{\rm mag}(t)]dt$.

\subsection{Properties of injected particles}
\subsubsection{Energy injection}
Both rotational energy injection \citep{Dai2017,KM2017} and magnetic energy injection \citep{Beloborodov2017,MM2018,ZW2021} models have been proposed to explain the PRS luminosity as well as the large RM detected for FRB 20121102A~\citep{Michilli2018}. 
The observed persistent radio emission from FRBs can be interpreted as synchrotron emission from electron-positron pairs in the nebula. 

The upper limit of the energy stored in the nebula is
\begin{eqnarray}
\label{Einj_max}
    {\rm Max}(\mathcal{E}^{\rm inj}_{e}) \approx \epsilon_e \mathcal{N}_{\rm sd}(t_{\rm age})\int_0^{t_{\rm age}} L_{\rm sd}(t^\prime)dt^\prime \nonumber \\ + \  \epsilon_e \mathcal{N}_{\rm mag}(t_{\rm age})\int_0^{t_{\rm age}} L_{\rm mag}(t^\prime)dt^\prime.
\end{eqnarray}
Here $\epsilon_e \approx 1-\epsilon_B$ is the injection efficiency, $\mathcal{N}_{\rm sd/mag}(t_{\rm age})=1$ for $t_{\rm age}<t_{\rm sd/mag}$ and $\mathcal{N}_{\rm sd/mag}(t_{\rm age})=(t_{\rm sd/mag}/t_{\rm age})[1 + {\rm log}(t_{\rm age}/t_{\rm sd/mag})$ for $t_{\rm age}>t_{\rm sd/mag}$, accounts for the effect of adiabatic energy loss from the magnetar wind nebula which is especially significant at late times. As radiative energy losses are also relevant, equation~(\ref{Einj_max}) provides a strict upper limit for energy contained in the nebula.

\subsubsection{Particle number distribution}
The rotational or magnetic energy is injected into the nebula together with the particles. The injected particles are accelerated to relativistic energies inside or around the termination shock before entering the nebula. The lepton injection rate $\dot{n}_{\mathcal{E}_e}^{\rm inj}$ is given by a broken power-law function (see e.g, \citealt{TT2010,Murase2015})
\begin{equation}
\label{n_e_inj}
    \mathcal{E}_e^2 \dot{n}_{\mathcal{E}_e}^{\rm inj} = \frac{3\epsilon_e (L_{\rm sd} + L_{\rm mag})}{4\pi R_{\rm nb}^2 c \mathcal{R}_0} \left\{
\begin{array}{ll}
(\gamma_e/\gamma_b)^{2-q_1}, & \gamma_e \leq \gamma_b \vspace{0.2cm} \\
(\gamma_e/\gamma_b)^{2-q_2}, & \gamma_b < \gamma_e \\
\end{array}
\right.
\end{equation}
where $R_{\rm nb}$ is the nebula radius, $\mathcal{R}_0 \sim (2-q_1)^{-1} + (q_2-2)^{-1} \sim 5$ is the correction factor for the lepton normalization, $q_1 < 2$ and $q_2 \geq 2$ are the low- and high-energy spectral indices. For our analysis, we adopt the characteristic LF of the accelerated leptons to be $\gamma_b \sim 10^{5}$ for rotation-powered and $\gamma_b \sim 10^{3}$ for magnetic-powered models. The spectral indices are $q_1=1.5$ and $q_2=2.5$ for rotation-powered scenario, and $q_1=q_2=2.0$ for magnetic-powered case. 

We assume $10^3 \leq \gamma_e \leq 10^7$, based on the multi-wavelength modelling of young Galactic PWNe \citep{TT2013}. The effective pair multiplicity can be generally expressed in terms of $\gamma_b$, $q_1$ and $q_2$ (see \citealt{Murase2015}). 
The characteristic Lorentz factor of accelerated electrons and positrons is adopted within the typical range inferred from relativistic shock and reconnection simulations. This choice is intended as an empirical representation of the injection process, rather than a direct outcome of detailed magnetohydrodynamic (MHD) evolution. Consequently, our formulation captures the broad energetics of the post-shock nebula without explicitly solving the forward–reverse shock dynamics for arbitrary magnetization.

\section{Properties of the compact PRS}
\label{Sec3}
Due to energy injection into the young magnetar, the wind nebula and SN ejecta tend to evolve together. Here we discuss their combined evolution and compute the synchrotron emission from the magnetised nebula that leads to the observed persistent radio counterparts for FRB sources.

\subsection{MWN and SNR evolution}
The density profile of the SN ejecta can be described using a power-law function $\rho_{\rm ej} = (3-\delta) M_{\rm ej}/4\pi R_{\rm ej}^3 (R/R_{\rm ej})^{-\delta}$ \citep{CS1989,KB2010}, where $\delta=1$ is the fiducial value of the index, $M_{\rm ej}$ is the ejecta mass and $R_{\rm ej}$ is the ejecta radius. The time evolution of the SN ejecta internal energy is given by
\begin{equation}
    \frac{d\mathcal{E}_{\rm int}}{dt} = \epsilon_e f_{\rm dep,sd/mag}L_{\rm sd/mag} + f_{\rm dep,rd}L_{\rm rd} - L_{\rm sn} - \frac{\mathcal{E}_{\rm int}}{R_{\rm ej}}\frac{dR_{\rm ej}}{dt}
\end{equation}
where $f_{\rm dep,sd/mag/rd}$ is the energy fraction deposited from NS spindown/magnetic field injection/radioactive decay estimated from \citet{Kashiyama2016}, $L_{\rm rd}$ is the radioactive decay power, $L_{\rm sn}$ is the SN luminosity and the last term represents energy loss due to adiabatic expansion.

\subsection{Synchrotron emission}
We estimate the radii of the ejecta and nebula using (see also, \citealt{Metzger2014,Kashiyama2016})
\begin{eqnarray}
\label{R_nb}
    \frac{dR_{\rm nb}}{dt} &=& \sqrt{\frac{7}{6(3-\delta)}\frac{\mathcal{E}_{\rm tot}}{M_{\rm ej}}\left(\frac{R_{\rm nb}}{R_{\rm ej}}\right)^{3-\delta}} + \frac{R_{\rm nb}}{t},\\
    \label{R_ej}
    \frac{dR_{\rm ej}}{dt} &=& V_{\rm ej},
\end{eqnarray}
where the first term on the right-hand side of equation~(\ref{R_nb}) corresponds to the nebula velocity in the ejecta rest frame and $V_{\rm ej}$ is the ejecta velocity. The initial ejecta velocity is almost constant, $V_{\rm ej,i} \approx 10^4\, {\rm km/s}\, (\mathcal{E}_{\rm sn}/10^{51}\, {\rm erg})^{1/2} (M_{\rm ej}/M_{\odot})^{-1/2}$, in the Sedov-Taylor expansion phase prior to energy injection. If $R_{\rm nb} > R_{\rm ej}$, the ejecta and nebula radii evolve together with the same velocity $dR_{\rm ej}/dt = dR_{\rm nb}/dt = V_{\rm ej,f} = \sqrt{2(\mathcal{E}_{\rm tot}+\mathcal{E}_{\rm SN})/M_{\rm ej}}$ as the injected energy significantly accelerates the ejecta through the magnetised wind. 

The average magnetic field in the nebula is given by $B_{\rm nb} = \sqrt{6\mathcal{E}_B/R_{\rm nb}^3}$, where $\mathcal{E}_B$ is the magnetic energy in the nebula. The magnetic energy in the nebula evolves as (see e.g., \citealt{Murase2016,Murase2021})
\begin{equation}
    \frac{d\mathcal{E}_B}{dt} = \epsilon_B(L_{\rm sd}+L_{\rm mag}) - c_B \frac{\mathcal{E}_B}{R_{\rm nb}} \frac{dR_{\rm nb}}{dt},
\end{equation}
In this work, we consider $c_B=0$ in the limit when energy loss due to adiabatic expansion is negligible, as done previously for Galactic PWNe \citep{TT2010} and pulsar-powered PWNe \citep{Murase2015,Murase2016}.

Here the post-shock energy densities are parameterized using $\epsilon_e$ and $\epsilon_B$, which describe the downstream energy partition between relativistic particles and magnetic fields, respectively. These quantities are treated phenomenologically rather than derived self-consistently from the upstream wind magnetization $\sigma$. Implicit in our approach is that a fraction $0 < \eta_{\rm diss} < 1$ of the upstream Poynting flux is efficiently converted into the downstream internal energy of particles. We assume that $\eta_{\rm diss}$ is sufficiently large for the downstream energy fractions ($\epsilon_B$, $\epsilon_e$) to be physically reasonable, implying that a significant portion of the Poynting flux is locally dissipated. This can be achieved through various mechanisms, including magnetic reconnection in a striped wind~\citep{Coroniti1990,Cerutti2020}, dissipation at termination shock~\citep{KC1984}, current-driven kink instabilities~\citep{Mizuno2011}, and turbulence in the nebula~\citep{Porth2014,ZrakeArons2017}. As the efficiencies of these processes remain uncertain and model-dependent, a detailed treatment of the dissipation physics is beyond the scope of this work.

We compute the synchrotron emission from the nebula by solving kinetic equations for the photons and electrons and positrons (see e.g., \citealt{Murase2021})
\begin{eqnarray}
\label{kin_eqs}
    \frac{\partial n_{\mathcal{E}_{\gamma}}}{\partial t} &=& -n_{\mathcal{E}_\gamma} \left(\frac{1}{t_{\rm comp}^{\rm nb}} + \frac{1}{t_{\rm esc}^{\rm nb}} + \frac{1}{t_{\gamma\gamma}}\right) + \frac{\partial}{\partial t}\left(n_{\mathcal{E}_\gamma}^{\rm IC} + n_{\mathcal{E}_\gamma}^{\rm syn}\right),\nonumber \\
    \frac{\partial n_{\mathcal{E}_{e}}}{\partial t} &=& -\frac{\partial}{\partial \mathcal{E}_e}[(P_{\rm ad} + P_{\rm syn} + P_{\rm IC})n_{\mathcal{E}_e}] + \frac{\partial n_{\mathcal{E}_e}^{\gamma \gamma}}{\partial t}  + \dot{n}_{\mathcal{E}_e}^{\rm inj} 
\end{eqnarray}
where $t_{\rm comp}^{\rm nb}$ is the Comptonisation timescale in the nebula, $t_{\rm esc}^{\rm nb}=R_{\rm nb}/c$ is the photon escape timescale, $t_{\gamma \gamma}$ is the photon annihilation timescale, $\partial n_{\mathcal{E}_{\gamma}}^{\rm IC/syn}/\partial t$ is the photon generation rate from IC/synchrotron process. $P_{\rm ad}$, $P_{\rm syn}$ and $P_{\rm IC}$ are the energy-loss rates due to adiabatic expansion, synchrotron radiation and IC, respectively, $\partial n_{\mathcal{E}_{e}}^{\gamma \gamma}/\partial t$ is the electron-positron production rate from photon annihilation and $\dot{n}_{\mathcal{E}_e}^{\rm inj}$ is the electron injection rate from equation~(\ref{n_e_inj}).

\section{Physical conditions} 
\label{Sec4}
Here we examine the physical conditions that are imposed on the parameters of the magnetised nebula and SN ejecta which we derived earlier. These conditions are necessary to qualitatively explain the observations for the persistent radio counterparts associated with FRBs. 

\subsection{Constraints on model parameters} \label{Sec4.1}
We list here the constraints on the model parameters: the dipolar magnetic field $B_{\rm dip}$, the initial spin period $P_i$, NS age $t_{\rm age}$, SN ejecta mass $M_{\rm ej}$ and explosion energy $\mathcal{E}_{\rm SN}$. 

\begin{figure*}
\includegraphics[width=0.32\textwidth]{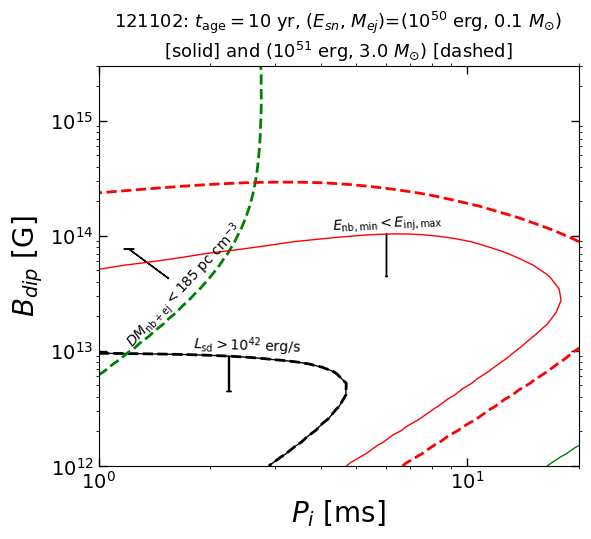} 
\includegraphics[width=0.32\textwidth]{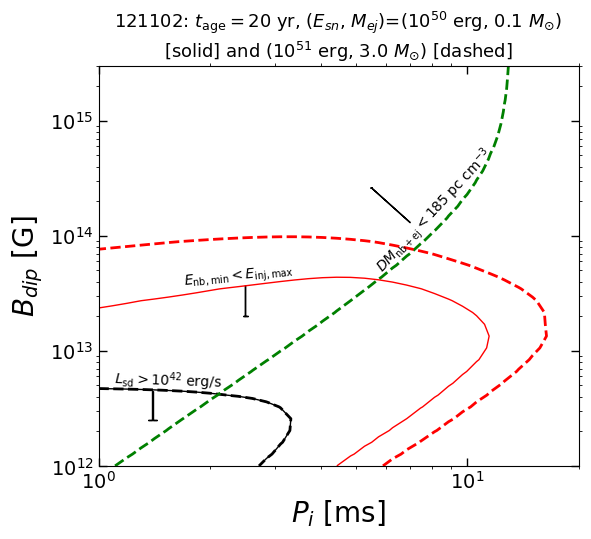}
\includegraphics[width=0.32\textwidth]{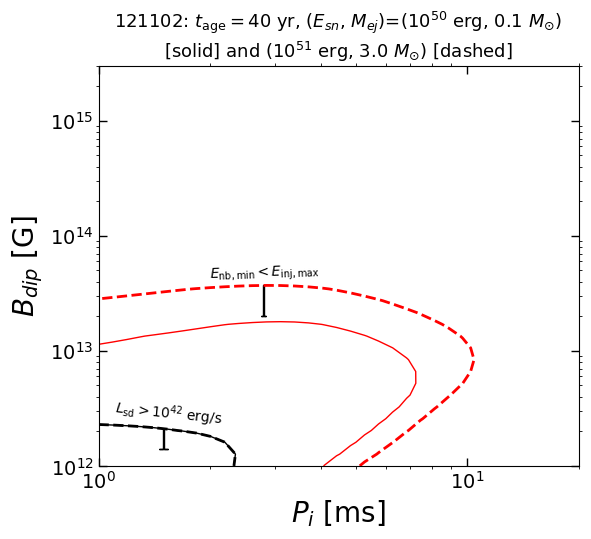}\\
\includegraphics[width=0.32\textwidth]
{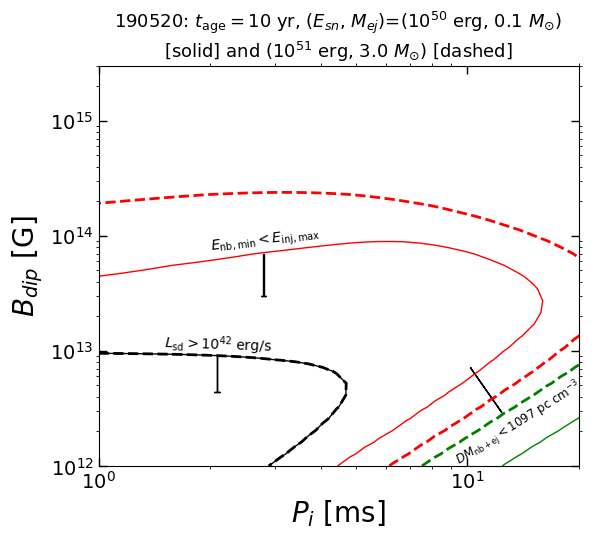}
\includegraphics[width=0.32\textwidth]{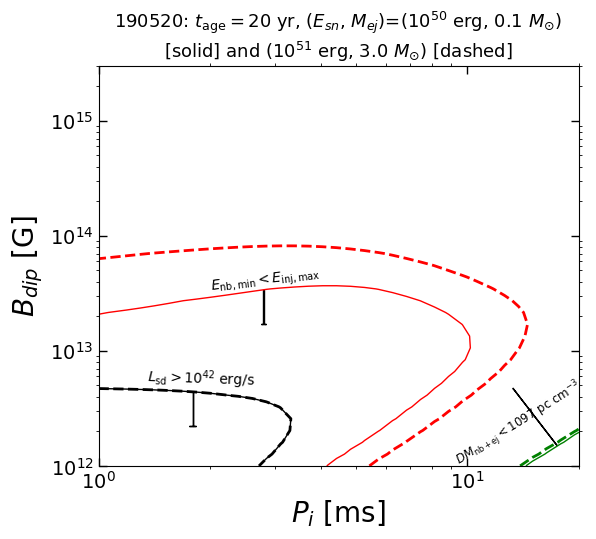}
\includegraphics[width=0.32\textwidth]{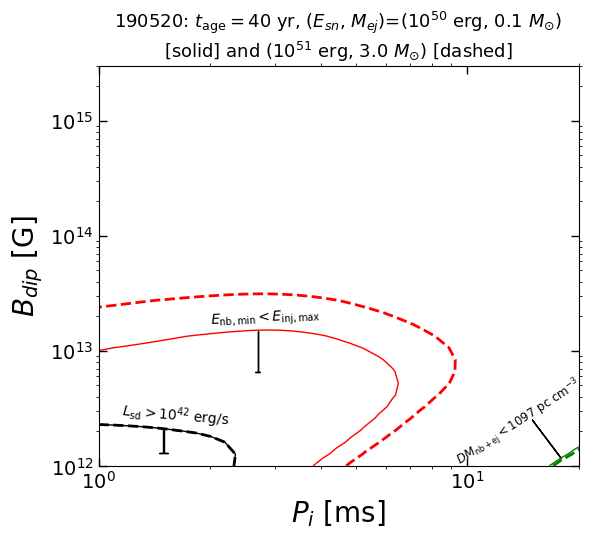}\\
\includegraphics[width=0.32\textwidth]
{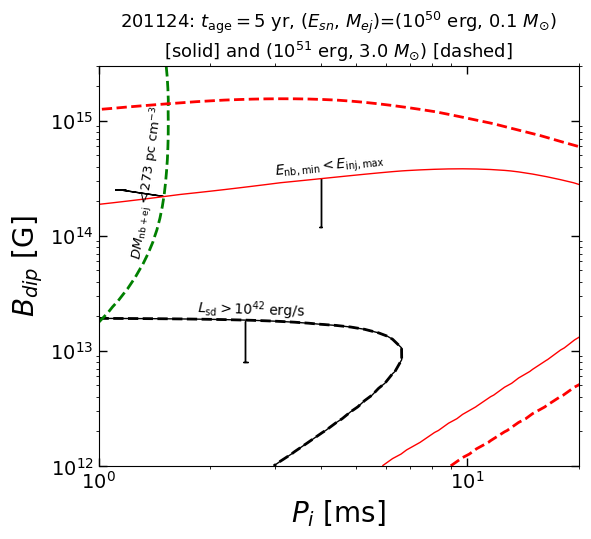}
\includegraphics[width=0.32\textwidth]{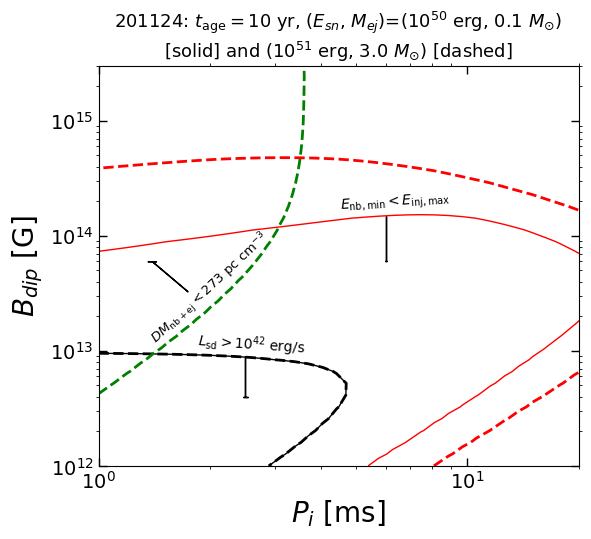}
\includegraphics[width=0.32\textwidth]{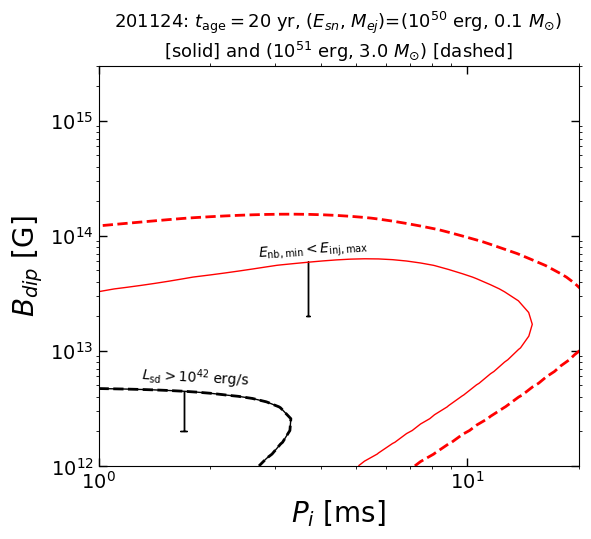}
\caption{Constraints on NS parameters $B_{\rm dip}$ and $P_i$ from the nebula energy requirement (red curves), ejecta and nebula contribution to the source DM (green curves), and the NS spindown luminosity (black curves) are shown for FRB 20121102A/20190520B/20201124A in top/middle/bottom row. In each panel, the solid curves show the results for ($E_{\rm sn}$, $M_{\rm ej}$) = ($10^{50}\,{\rm erg}$, $0.1\,M_{\odot}$) while the dashed curves show results for ($E_{\rm sn}$, $M_{\rm ej}$) = ($10^{51}\,{\rm erg}$, $3.0\,M_{\odot}$). For the magnetic energy injection, we fix $B_{\rm int} = 10^{16}\,{\rm G}$ and $t_{\rm inj} = 0.6\,{\rm yr}$ for all cases. For FRBs 20121102A and 20190520B, we vary $t_{\rm age} = 10\,{\rm yr}$ (left panel), $20\,{\rm yr}$ (middle panel) and $40\,{\rm yr}$ (right panel) as $t_{\rm obs}>5\,{\rm yr}$. In case of FRB 20201124A, results are shown for $t_{\rm age} = 5\,{\rm yr}$ (left panel), $10\,{\rm yr}$ (middle panel) and $20\,{\rm yr}$ (right panel).
} 
\label{FRB_constraint_plots}
\end{figure*}

\begin{itemize}
    \item \emph{Magnetised nebula energy requirement:} The minimum energy in relativistic electrons and positrons required to explain the quasi-steady radio emission from FRB sources should be less than the maximum energy stored in the nebula which is given by ${\rm Max}(\mathcal{E}_e^{\rm inj})$ (see e.g., equation~\ref{Einj_max}).
    The observed quasi-steady radio spectrum for FRBs can be fitted by
    \begin{equation}
    \label{qs_radiospec}
        \nu F_\nu \approx \frac{\mathcal{L}_{\rm nb}}{4\pi d_L^2}\left\{
\begin{array}{ll}
(\nu/\overline{\nu})^{p_1}, & \nu < \overline{\nu} \vspace{0.2cm} \\
(\nu/\overline{\nu})^{p_2}, & \nu \geq \overline{\nu} \\
\end{array}
\right.
    \end{equation} 
    Here $\overline{\nu} \sim 10-30\,{\rm GHz}$, 
    $\mathcal{L}_{\rm nb} \sim 10^{38}-10^{39}\,{\rm erg/s}$, $p_1 \sim 0.6-2.0$ and $p_2 \sim -0.5$ for the five FRBs currently detected with a confirmed PRS~\citep{Chatterjee2017,Niu2022,Bruni2023,Bruni2024,Moroianu2025}. The luminosity distance $d_L \sim 400-1200\,{\rm Mpc}$ is determined by the known redshift $z \sim 0.1-0.2$.  
    
    We assume that the bolometric luminosity for nebula $\mathcal{L}_{\rm nb}$ peaks in the radio bands, to estimate the minimum required energy ${\rm Min}(\mathcal{E}_e^{\rm nb})$ stored in the electrons and positrons. The LF for radio-emitting electron is $\gamma_e \approx (4\pi m_e c \overline{\nu}/3e B_{\rm nb})^{1/2}$ and the energy loss rate is $P_e \approx (4/3)\sigma_T c \gamma_e^2 B_{\rm nb}^2/8\pi$. The total number of electrons and positrons in the magnetised nebula is roughly $N_e \approx \mathcal{L}_{\rm nb}/P_e$ and the minimum required energy in the nebula is then (see \citealt{KM2017})
    \begin{equation}
        {\rm Min}(\mathcal{E}_{e}^{\rm nb}) = N_e \gamma_e m_e c^2 \approx \frac{3\sqrt{3}(\pi e m_e c)^{1/2} \mathcal{L_{\rm nb}}}{\sigma_T \overline{\nu}^{1/2}B_{\rm nb}^{3/2}}
    \end{equation}
    Therefore, the rotational and magnetic energy injection should be large enough to provide sufficient energy to the nebula. Also, the NS should be young enough such that adiabatic energy losses are not significant.\\ 
    \item \emph{Constraint on source DM:} The number density of free electrons in the SN ejecta and magnetised nebula are given by $n_{\rm e,ej} \approx 3M_{\rm ej}/(4\pi R_{\rm ej}^3 \mu_e \overline{A} m_{\rm H})$ and $n_{\rm e,nb} \approx 3M_{\rm nb}/(4\pi R_{\rm nb}^3 \mu_e \overline{A} m_{\rm H})$, respectively. Here $M_{\rm ej/nb}$ is the ejecta/nebula mass, $\overline{A}=10$ is the mean atomic mass number and $\mu_e \approx 1$ for the singly ionized state corresponding to electron temperature $\mathcal{T}_e \sim 10^4\, {\rm K}$ (see \citealt{KM2017}). The total DM contribution close to the source, from the SN ejecta and magnetised wind nebula is
    \begin{equation}
        {\rm DM}_{\rm ns} \approx {\rm DM}_{\rm ej}+{\rm DM}_{\rm nb} \approx n_{\rm e,ej}R_{\rm ej} + n_{\rm e,nb}R_{\rm nb},
    \end{equation}
    where $R_{\rm nb}$ and $R_{\rm ej}$ are obtained by solving equations~(\ref{R_nb}) and~(\ref{R_ej}). Although $M_{\rm nb} \gtrsim 10^{-9}-10^{-5}\,M_\odot$ is largely uncertain~\citep{Murase2015}, a reasonable lower limit can be obtained using the Goldreich-Julian (GJ) density~\citep{GJ1969}
    \begin{equation}
        M_{\rm nb}/t_{\rm age} \gtrsim \dot{M}_{\rm \pm} \approx (2.4\times10^{-11}\, M_{\odot}/{\rm s})\, \mu_{\pm,6}B_{\rm int,15}P_{i,-3}^{-2},
    \end{equation}
    where $\dot{M}_{\rm \pm}$ is the mass-loss rate due to electron-positron pairs and $\mu_{\pm} \sim 10^{5-6}$ is the pair multiplicity defined as the ratio of the pair density to the Goldreich-Julian (GJ) density. It is difficult to infer a plausible theoretical value for the pair multiplicity from first principles. In this study, we adopt $M_{\rm nb} \sim 10^{-7}\,M_\odot$ as a fiducial value~\citep{Murase2015}. The ejecta DM contribution from the SNR should evolve over time as ${\rm DM}_{\rm ej} \propto t^{-2}$. The NS needs to be old enough such that ${\rm DM}_{\rm ns} \approx n_{\rm e,ej}R_{\rm ej} + n_{\rm e,nb}R_{\rm nb} \lesssim {\rm DM}_{\rm host}$.\\
    \item \emph{Non-attenuation of radio signal:} Radio pulses that are produced in the NS magnetosphere can be diminished either through scattering or absorption process in the SN ejecta and magnetised nebula. Free-free absorption in the SN ejecta is one of the relevant processes and the NS should be old enough such that the opacity at $\sim$ GHz frequencies (see e.g., \citealt{Murase2017})
    \begin{equation}
        \tau_{\rm ff}\approx 2.1\times10^{-25}\, \mathcal{T}_{e,4}^{-1.35} \int dr n_{\rm e,ej} n_{\rm i,ej} \overline{Z}^2,
    \end{equation}
does not exceed unity. Here $\overline{Z} \sim \overline{A}/2$ and $n_{\rm i,ej}$ is the number density of ions in the ejecta. Furthermore, it is expected that synchrotron self-absorption (SSA) is subdominant in magnetised nebula indicating that \citep{Yang2016}
    \begin{equation}
        \tau_{\rm sa} = R_{\rm nb} \int \mathcal{E}_e \frac{dn_{\mathcal{E}_e}}
    {d\mathcal{E}_e} \sigma_{\rm sa}(\nu, \mathcal{E}_e)
    \end{equation}
is also less than unity, for a energy-dependent SSA cross section $\sigma_{\rm sa}$ \citep{GS1991}.\\
    \item \emph{Size constraint for FRB source:} Lastly, the size of the magnetised nebula should not exceed the observed upper limit on the size of the PRS, given by its angular size and luminosity distance for the FRB. From VLBI and VLA observations of FRBs 20121102A, 20190520B and 20201124A, $R_{\rm nb} \lesssim 1-10\, {\rm pc}$ (see \citealt{Tendulkar2017,Niu2022,Bruni2023}).
\end{itemize}

\subsection{Implications for FRBs detected with PRS}

To compute the synchrotron emission from magnetized nebula, we first need to constrain the NS parameters including $B_{\rm dip}\sim 10^{12}-10^{15}\,{\rm G}$, $P_i\sim1-30\,{\rm ms}$ and $t_{\rm age} \gtrsim t_{\rm obs}$, where $t_{\rm obs}$ is the time since initial detection of the FRB source. For our analysis, we consider two progenitors: (a) an ultra-stripped SNe with $M_{\rm ej}\sim0.1\,M_{\odot}$ and $\mathcal{E}_{\rm SN}\sim10^{50}\,{\rm erg}$ (USSN model), and (b) a conventional core-collapse SNe with $M_{\rm ej}\sim3.0\,M_{\odot}$ and $\mathcal{E}_{\rm SN}\sim10^{51}\,{\rm erg}$ (CCSN model). For the injected magnetic energy (see equation~\ref{Lmag}), we adopt $B_{\rm int}=10^{16}\,{\rm G}$ and $t_{\rm inj}=0.6\,{\rm yr}$ as in \citet{MM2018}. We discuss the combination of NS parameters $B_{\rm dip}$, $P_i$ and $t_{\rm age}$ for which the magnetar satisfies all the necessary constraints as discussed in Section~\ref{Sec4.1}.  

Figure~\ref{FRB_constraint_plots} shows the constraints on $P_i -B_{\rm dip}$ parameter space for the nebula energy requirement in red, DM contribution from the ejecta and nebula in green, and the NS spindown luminosity in black. The results for FRBs 20121102A, 20190520B and 20201124A are shown in top, middle and bottom row panels, respectively, with solid/dashed curves corresponding to USSN/CCSN model in each panel. Along with NS spindown luminosity, we also consider magnetic energy injection with $B_{\rm int}=10^{16}\,{\rm G}$ and $t_{\rm inj}=0.6\,{\rm yr}$ for all cases. We find that the DM criteria for SN ejecta and MWN contribution (${\rm DM}_{\rm ej+nb} < {\rm DM}_{\rm host}$), and the MWN minimum energy requirement ($\varepsilon_{\rm nb,min}<\varepsilon_{\rm inj,max}$) are the most constraining. In the left, center and right column panels, we vary the NS age: $t_{\rm age}\sim 10-40\,{\rm yr}$ for FRBs 20121102A and 20190520B, and $t_{\rm age}\sim 5-20\,{\rm yr}$ for FRB 20201124A. 
With an increase in $t_{\rm age}$, $DM_{\rm nb+ej}$ decreases faster due to expansion as compared to the reduction in $E_{\rm inj, max}$ from adiabatic energy losses, thereby allowing a larger $P_i-B_{\rm dip}$ parameter space. For a given $P_i$, ${\rm DM}_{\rm ej+nb}$ reduces faster for a larger $B_{\rm dip}$, as more energy gets injected into the SN ejecta within a shorter timescale leading to a faster ejecta expansion. However, a smaller value of $B_{\rm dip}$ is allowed for a given $P_i$ to prevent considerable adiabatic losses. 

For a given $t_{\rm age}$, it is easier to satisfy the DM criterion (${\rm DM}_{\rm ej+nb} < {\rm DM}_{\rm host}$) in the order FRB 20190520B > FRB 20121102A > FRB 20201124A, which directly follows from the inferred $DM_{\rm host}$ values for these three localized sources. 
However, the energy criterion ($\varepsilon_{\rm nb,min}<\varepsilon_{\rm inj,max}$) is satisfied in a larger $P_i-B_{\rm dip}$ parameter space for FRB 20201124A as compared to the other two sources, primarily due to smaller $q_1$ leading to more energy injection near the peak of the spectrum (see equation~\ref{qs_radiospec}). 
Although the qualitative trends in parameter constraints obtained from near-source DM and nebula energy criteria are effectively same, for both USSN and CCSN progenitors at a given $t_{\rm age}$, the DM (energy) constraint is harder (easier) to satisfy for the CCSN case due to their larger $M_{\rm ej}$ ($E_{\rm sn})$. 
For $t_{\rm age} \gtrsim 10\,{\rm yr}$ and USSN (CCSN) progenitor, the allowed parameter space satisfying both these conditions is limited to: $P_i \sim 1-10\,{\rm ms}$ and $B_{\rm dip}\sim 10^{12}-10^{14}\,{\rm G}$ ($P_i \sim 1-2.5\,{\rm ms}$ and $B_{\rm dip}\sim 5\times10^{12}-2\times10^{14}\,{\rm G}$) for FRB 20121102A, 
$P_i \sim 1-10\,{\rm ms}$ and $B_{\rm dip}\sim 10^{12}-10^{14}\,{\rm G}$ ($P_i \sim 1-20\,{\rm ms}$ and $B_{\rm dip}\sim 10^{12}-2\times10^{14}\,{\rm G}$) for FRB 20190520B, and 
$P_i \sim 1-20\,{\rm ms}$ and $B_{\rm dip}\sim 10^{12}-10^{14}\,{\rm G}$ ($P_i \sim 1-3\,{\rm ms}$ and $B_{\rm dip}\sim 5\times10^{12}-3\times10^{14}\,{\rm G}$) for FRB 20201124A. 
For both USSN and CCSN cases, the parameter space satisfying $L_{\rm sd} \gtrsim 10^{42}\,{\rm erg/s}$ overlaps with that allowed by ${\rm DM}_{\rm ej+nb} < {\rm DM}_{\rm host}$ and  $\varepsilon_{\rm nb,min}<\varepsilon_{\rm inj,max}$, especially for $t_{\rm age} \gtrsim 10\,{\rm yr}$. This indicates that NS spindown luminosity is likely to be the primary energy source for these three FRBs.

\section{Observed FRB properties}
\label{Sec5}
With the constraints obtained for NS parameters ($P_i$, $B_{\rm dip}$, $t_{\rm age}$) in Section~\ref{Sec4}, we compute the spectral energy distribution (SED) and light curve of the associated PRS, in addition to the time evolution of source DM for the three FRBs to compare them with the radio data. Efficient conversion of rotational energy to particle energy in the termination shock region is required to explain the observed quasi-steady radio emission. We consider two scenarios: (a) rotation-powered model with ($\epsilon_B=0.01$, $\gamma_b=10^5$, $q_1=1.5$, $q_2=2.5$), and (b) magnetar-flare-powered model with ($\epsilon_B=0.1$, $\gamma_b=10^3$, $q_1=q_2=2$). We use an opacity of $\kappa=0.05\,{\rm cm^{2}/g}$ for the SN ejecta, and set the NS mass $M_{\rm NS}=1.4\,M_{\odot}$, radius $R_{\rm NS}=10\,{\rm km}$ and initial SN explosion radius $R_0=10^6\,{\rm km}$.

\begin{figure*}
\includegraphics[width=0.32\textwidth]{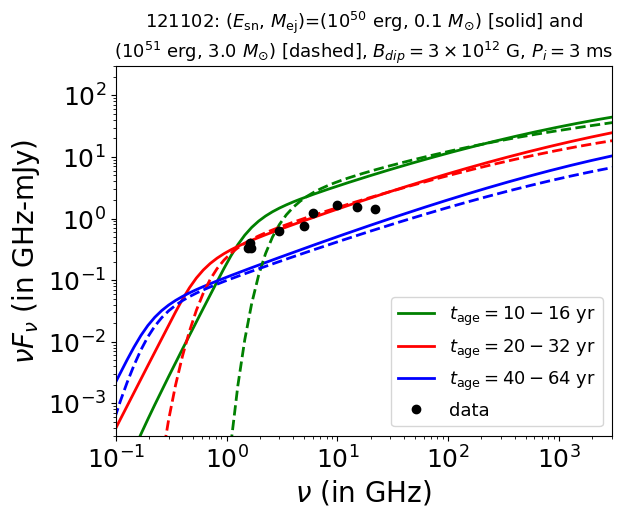}
\includegraphics[width=0.32\textwidth]{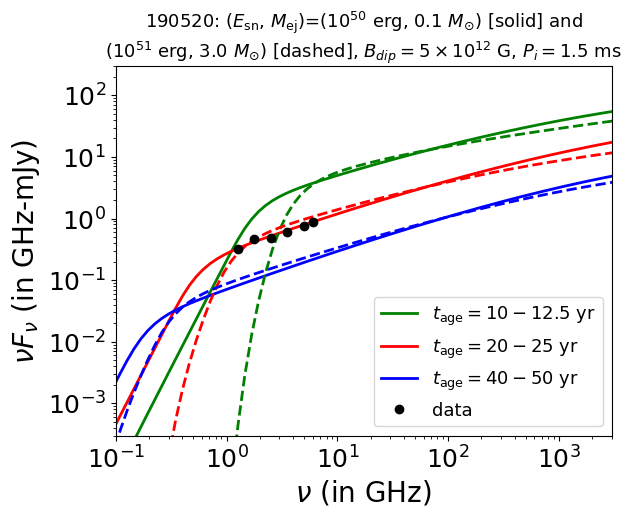}
\includegraphics[width=0.32\textwidth]{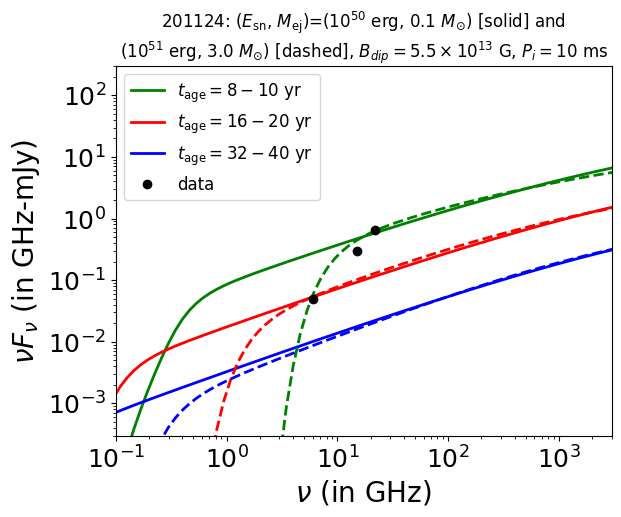}
\caption{Effect of NS age ($t_{\rm age}$) on the spectral energy distribution of PRS emission is shown for FRB 20121102A (20190520B) [20201124A] in the left (center) [right] panel. Data from radio observations of each source is shown with filled circles. In each panel, we show the results for USSN/CCSN progenitors using solid/dashed curves, for a fixed ($B_{\rm dip}$, $P_i$) combination and varying $t_{\rm age}$ -- including the best-fit NS age for the respective source. We assume the rotation-powered model with $\epsilon_B=0.01$, $\gamma_b=10^5$, $q_1=1.5$ and $q_2=2.5$.} 
\label{fig:SED_Mej_tage}
\end{figure*}

\begin{figure*}
\includegraphics[width=0.32\textwidth]{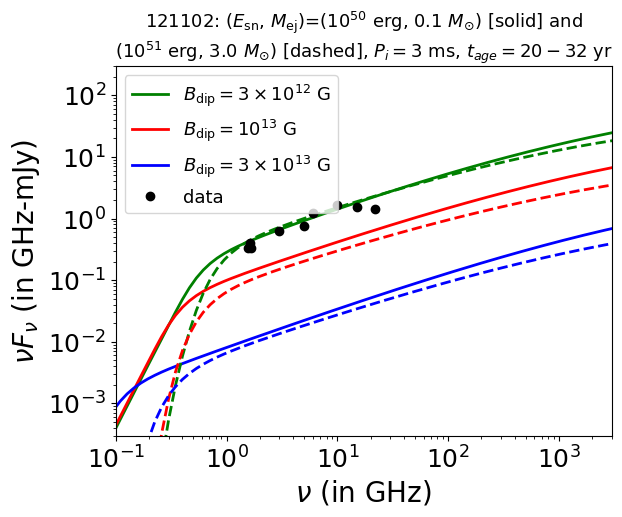}
\includegraphics[width=0.32\textwidth]{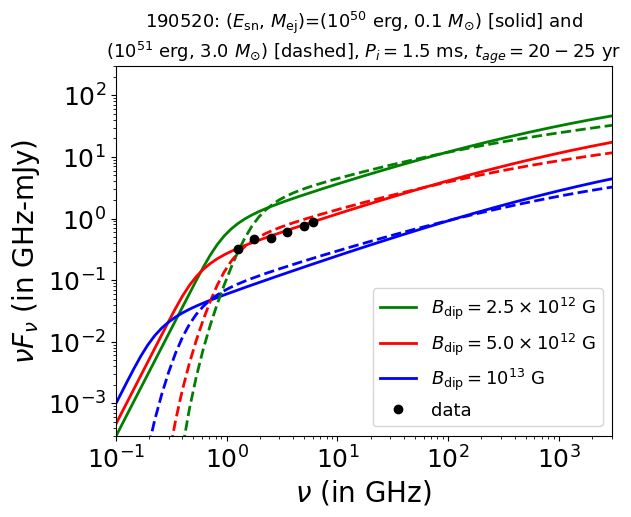}
\includegraphics[width=0.32\textwidth]{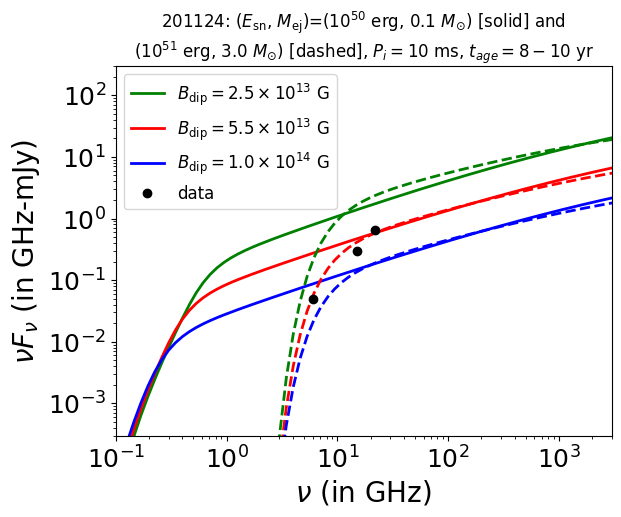}\\
\includegraphics[width=0.32\textwidth]{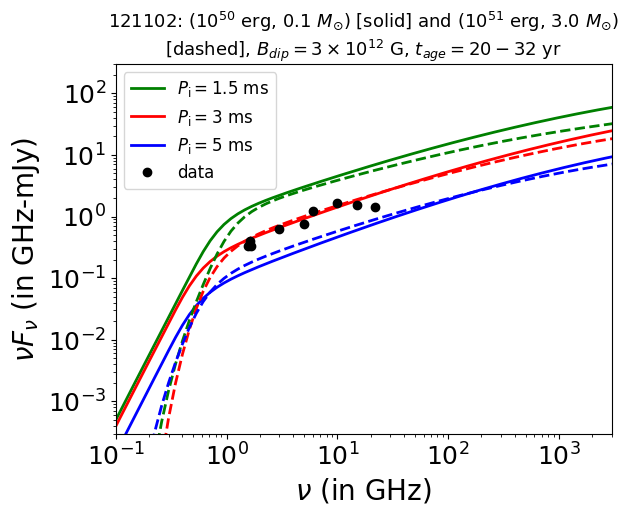}
\includegraphics[width=0.32\textwidth]{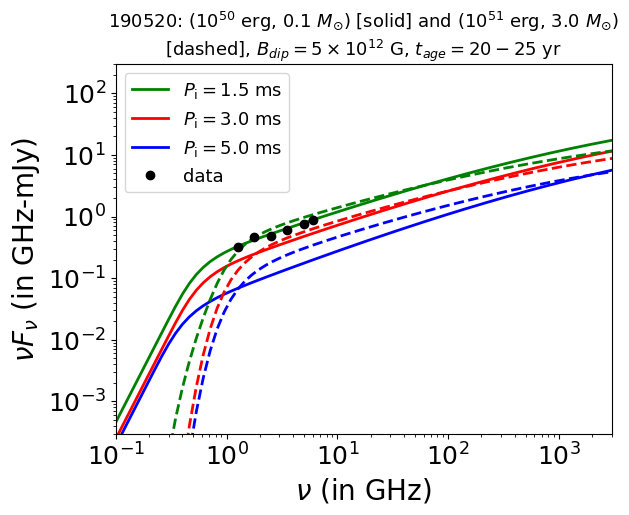}
\includegraphics[width=0.32\textwidth]{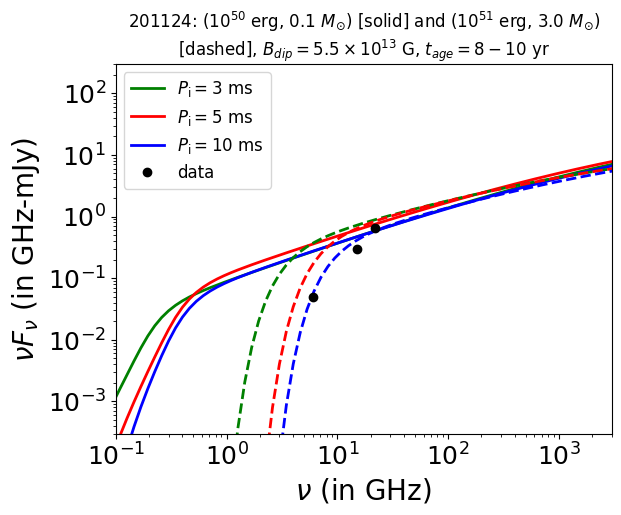}
\caption{Effect of NS dipolar magnetic field $B_{\rm dip}$ (in top row) and initial spin period $P_i$ (in bottom row) on the spectral energy distribution of PRS emission is shown for FRB 20121102A (20190520B) [20201124A] in the left (center) [right] column panels. In each panel, the data from radio observations is shown with filled circles and results for USSN/CCSN progenitors are shown using solid/dashed curves. For each FRB, we fix $t_{\rm age}$ to the best-fit value obtained from Figure~\ref{fig:SED_Mej_tage} and vary $B_{\rm dip}$ or $P_i$ -- including their best-fit combinations for the respective source. As in Figure~\ref{fig:SED_Mej_tage}, we assume microphysical parameters corresponding to the rotation-powered model.} 
\label{fig:SED_Mej_Bdip_Pi}
\end{figure*}

\subsection{SED \& light curve of the PRS}
We first numerically compute the radio SEDs for the nebular emission using the code developed by \citet{Murase2015,Murase2021}, which takes into account the effect of electron-positron pairs relevant for emission at $t \gtrsim t_{\rm sd}$. We incorporate the inverse Compton and synchrotron emission processes to solve the time-dependent kinetic equation~(\ref{kin_eqs}), and account for electromagnetic cascades. The external radiation fields are taken to be cosmic microwave background (CMB) and extragalactic background light (EBL). In our calculations, we have included the effect of synchrotron self-absorption (SSA) in magnetised nebula and free-free absorption in the SN ejecta, along with the Razin effect.

\begin{figure*}
\includegraphics[width=0.32\textwidth]{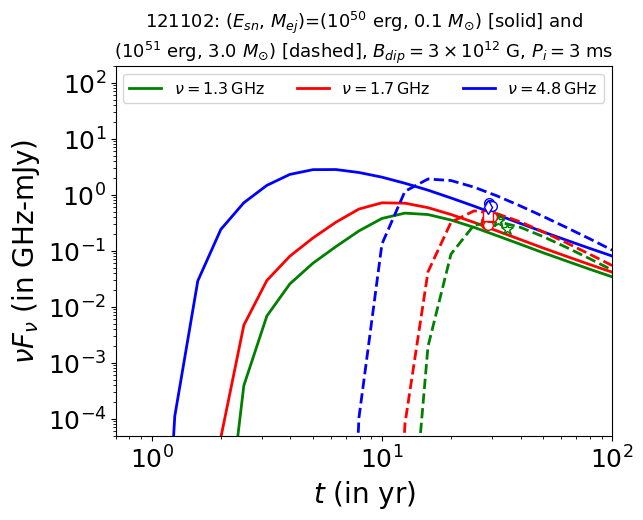}
\includegraphics[width=0.32\textwidth]{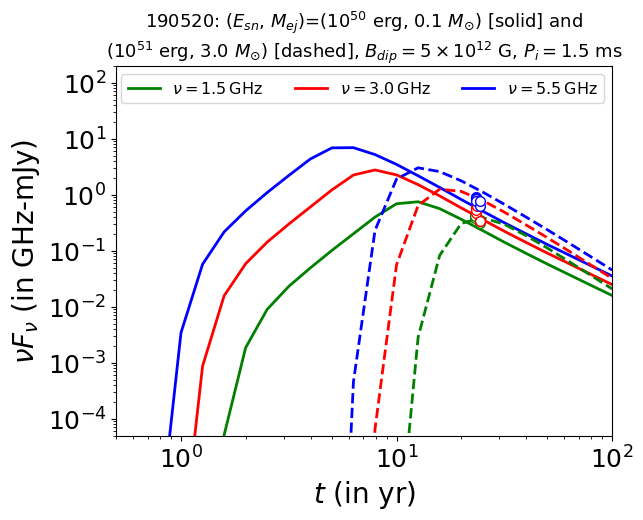}
\includegraphics[width=0.32\textwidth]{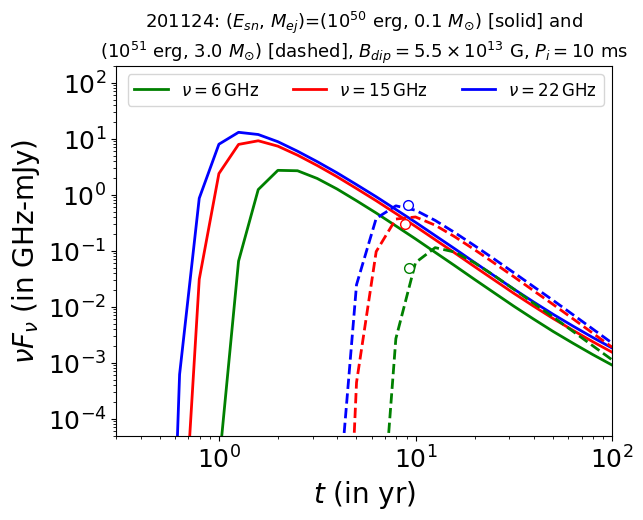}
\caption{Light curves for the persistent radio emission associated with FRB 20121102A (20190520B) [20201124A] are shown in the left (center) [right] panel. The corresponding data at various radio frequencies for each source is shown using unfilled circles. As earlier, the results for USSN and CCSN progenitors are shown with solid and dashed curves, respectively. We fix the NS parameters $B_{\rm dip}$, $P_i$ and $t_{\rm age}$ to their best-fit values obtained from Figures~\ref{fig:SED_Mej_tage} and~\ref{fig:SED_Mej_Bdip_Pi} for the rotation-powered model.} 
\label{fig:LC_Mej_Esn}
\end{figure*}

For the rotation-powered model and both progenitor types, the effect of $t_{\rm age}$ on the radio spectral energy distribution (SED) is shown for FRB 20121102A/20190520B/20201124A in the left/center/right panels of Figure~\ref{fig:SED_Mej_tage}, with the corresponding radio data shown with filled circles. As expected, the synchrotron flux is suppressed for larger NS age ($t_{\rm age} \gg t_{\rm sd}$) due to considerable adiabatic energy losses at late times. Based on comparison with the observed radio flux, we find a best-fit $t_{\rm age} \approx 20\,(20)\,[8]\,{\rm yr}$ for USSN and $t_{\rm age} \approx 32\,(25)\,[10]\,{\rm yr}$ for CCSN models, corresponding to FRB 20121102A (20190520B) [20201124A]. The low-energy synchrotron flux is highly suppressed due to SSA, especially for the large $M_{\rm ej}$ in case of CCSN progenitors. While the USSN model seems marginally more promising for both FRBs 20121102A and 20190520B to explain the observed PRS emission at smaller energies ($\nu \sim 1\,{\rm GHz}$), only CCSN model satisfactorily explains the observed emission for FRB 20201124A due to the large suppression of radio flux at smaller energies. 

Next we use the best-fit $t_{\rm age}$ obtained from Figure~\ref{fig:SED_Mej_tage} for each FRB to study the effect of NS parameters on the synchrotron radio flux. In particular, we analyze the effect of dipolar magnetic field $B_{\rm dip}$ on the radio SED as shown in the top-row panels of Figure~\ref{fig:SED_Mej_Bdip_Pi}. As in Figure~\ref{fig:SED_Mej_tage}, the results for FRB 20121102A/20190520B/20201124A are shown in the left/center/right column panels, with the same representation for progenitor types and radio data shown as filled circles. We find that dipolar field $B_{\rm dip}=3\times10^{12}\,(5\times10^{12})\,[5.5\times10^{13}]\,{\rm G}$ explains the observed radio flux for FRB 20121102A (20190520B) [20201124A]. With further increase in $B_{\rm dip}$, the NS spindown timescale ($t_{\rm sd} \propto B_{\rm dip}^{-2}P_i^2$) becomes smaller for a given $P_i$, which results in energy injection into the nebula at early times. This is then accompanied by significant adiabatic losses which leads to a reduced flux. 

The effect of NS spin period $P_i$ on the radio SED is shown in the bottom-row panels of Figure~\ref{fig:SED_Mej_Bdip_Pi}. As expected, the observed radio flux tends to be larger for a smaller $P_i$ as more rotational energy can be extracted from the NS. However, this difference is less noticeable for larger $B_{\rm dip}$ (like in case of FRB 20201124A), as $t_{\rm age} \gg t_{\rm sd}$ results in significant adiabatic energy loss. We find that $P_i=3\,(1.5)\,[10]\,{\rm ms}$ best explains the observed radio flux for FRB 20121102A (20190520B) [20201124A]. The low-energy radio flux is suppressed for both fixed $P_{\rm i}$ and fixed $B_{\rm dip}$ cases for the CCSN model, especially for FRB 20201124A, due to stronger SSA attenuation in the nebula. 
In summary, we find that a NS with $t_{\rm age}\approx 20\,{\rm yr}$, $B_{\rm dip}\approx (3-5)\times10^{12}\,{\rm G}$ and $P_i \approx 1.5-3\,{\rm ms}$ in a USSN progenitor can explain the observed PRS flux for FRBs 20121102A and 20190520B. However, in case of FRB 20201124A, the source is more likely to be a younger NS with $t_{\rm age}\approx 10\,{\rm yr}$, $B_{\rm dip}\approx 5.5\times10^{13}\,{\rm G}$ and $P_i \approx 10\,{\rm ms}$ in a CCSN progenitor. The properties of the central NS and its surrounding ejecta material are consistent with those deduced from previous studies (see e.g., \citealt{MM2018,ZW2021}).

\begin{figure*}
\includegraphics[width=0.32\textwidth]{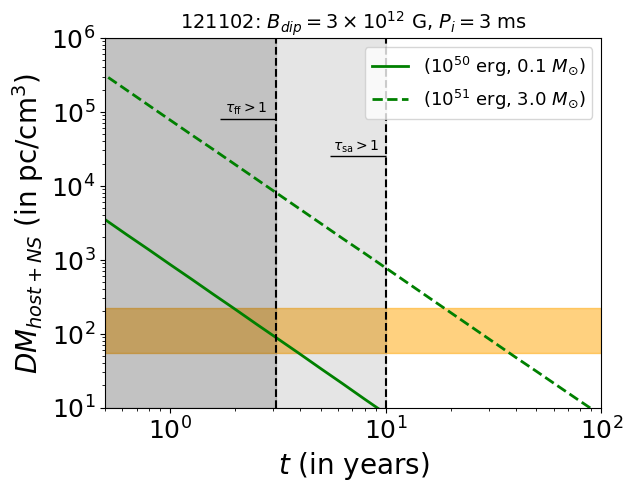}
\includegraphics[width=0.32\textwidth]{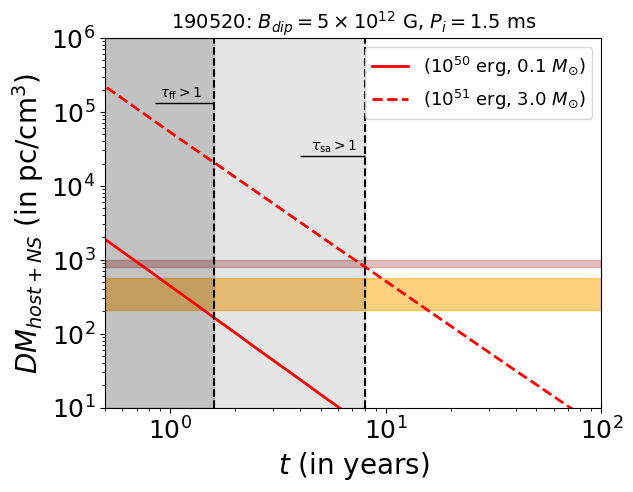}
\includegraphics[width=0.32\textwidth]{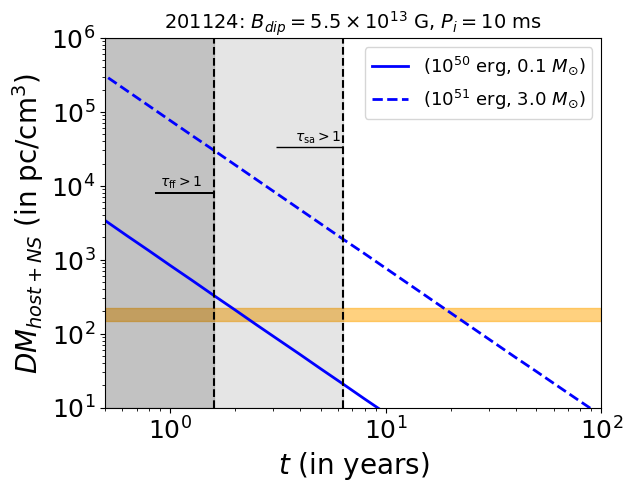}
\caption{Time evolution of ${\rm DM}_{\rm host+ns}$ is shown for FRB 20121102A (20190520B) [20201124A] in the left (center) [right] panel, by considering contribution from the expanding SN ejecta and MWN. Results for USSN and CCSN models are shown with solid and dashed curves, respectively, and the best-fit NS parameters ($B_{\rm dip}$, $P_i$) are the same as those in Figure~\ref{fig:LC_Mej_Esn}. The shaded yellow region in each panel denotes the host galaxy DM contribution including its uncertainty. For FRB 20190520B, we also show the previously estimated ${\rm DM}_{\rm host+ns}$, which likely includes contribution from foreground galaxy groups and clusters~\citep{Lee2023}. The dashed vertical lines (for the CCSN cases) correspond to minimum $t_{\rm age}$ obtained by imposing $\tau_{\rm sa} < 1$ in the MWN and $\tau_{\rm ff} < 1$ in the SN ejecta, respectively, with the grey shaded areas signifying the regions that are disallowed by the detectability of PRS.} 
\label{fig:DMevol_Mej_Esn}
\end{figure*}

We next compute the radio light curves associated with synchrotron emission from the magnetized nebula. To study the effect of progenitor model (USSN/CCSN), we use the best-fit $B_{\rm dip}$ and $P_i$ values obtained for each FRB source from Figure~\ref{fig:SED_Mej_Bdip_Pi}. The light curve data at various frequencies is obtained from: \citet{Marcote2017,Plavin2022,Rhodes2023} for FRB 20121102A, \citet{Niu2022,Zhang2023} for FRB 20190520B, and \citet{Bruni2023} for FRB 20201124A. The corresponding NS age is set to $t_{\rm age}=20\,(20)\,[10]\,{\rm yr}$ from Figure~\ref{fig:SED_Mej_tage}. We find that the radio flux predictions from the USSN and CCSN models, for the best-fit NS parameters, are effectively indistinguishable based on current observations for both FRBs 20121102A and 20190520B. However, in case of FRB 20201124A, recent observations at $\nu=6\,{\rm GHz}$ clearly favor the CCSN progenitor model, which also agrees with our results based on the radio SEDs shown in Figure~\ref{fig:SED_Mej_Bdip_Pi}. As expected, the radio light curve peaks at a later time and with a smaller synchrotron peak for larger ejecta mass $M_{\rm ej}$ and/or larger SN explosion energy $\mathcal{E}_{\rm sn}$ for the CCSN model. Although the synchrotron flux can reduce with an increase in $M_{\rm ej}$, especially for smaller $t_{\rm age}\sim 10\,{\rm yr}$ as in the case of FRB 20201124A, the spectrum is generally softer due to a smaller synchrotron peak. Lastly, the variation in synchrotron peak flux is marginally higher for CCSN progenitors with a larger SN explosion energy $\mathcal{E}_{\rm sn}$.

\subsection{NS age constraint from DM evolution}
For a FRB source located at redshift $z$, the observed DM can be separated into four primary components,
\begin{equation}
\label{DM_comps}
    {\rm DM}_{\rm obs} = {\rm DM}_{\rm MW} + {\rm DM}_{\rm halo} + {\rm DM}_{\rm IGM} + \frac{{\rm DM}_{\rm host}+{\rm DM}_{\rm ns}}{(1+z)}
\end{equation}
where ${\rm DM}_{\rm MW/halo}$ is the contribution from the Milky Way interstellar medium (ISM)/halo, ${\rm DM}_{\rm IGM}$ is the contribution from the intergalactic medium (IGM) and ${\rm DM}_{\rm host}$ is the contribution from the FRB host galaxy including its halo. Here ${\rm DM}_{\rm ns}$ only includes contributions from the MWN and SN ejecta that are near source. While the long-term DM variation is due to the expanding SNR \citep{YZ2017,PG2018}, random fluctuations in the DM can be caused by turbulent motions of filament \citep{Katz2021}. 

From theoretical and data-driven estimates of ${\rm DM}_{\rm MW/halo}$ and ${\rm DM}_{\rm IGM}$, assuming models for the average electron number density in these media (see e.g., \citealt{CL2002,CL2003,Planck2016,Prochaska2019,Platts2020}), the inferred ${\rm DM}_{\rm host}+{\rm DM}_{\rm ns}$ in equation~(\ref{DM_comps}) indicates significant contributions from the host galaxy disk, circumgalactic medium and near-source medium for most FRBs that have an associated PRS (see e.g., \citealt{Law2022}). In fact, from previous studies \citep{Chatterjee2017,Niu2022,Bruni2023,Lee2023}, we know that the host galaxy and near-source contribution ${\rm DM}_{\rm host}+{\rm DM}_{\rm ns} \sim 140\,(430)\,[185]\,{\rm pc\,cm^{-3}}$ for FRB 20121102A (20190520B) [20201124A] can be considerable. Using Illustris simulations to model the electron number density along various lines of sight, \citet{Zhang2020} showed that the typical DM contribution from host galaxies at $z \sim 0.2$, that are similar to the three FRBs considered here, can be $\sim 50-250\,{\rm pc\,cm^{-3}}$.

\begin{table*}
\begin{center}
\caption{The best fit NS parameters (dipolar magnetic field $B_{\rm dip}$, initial spin period $P_i$, age $t_{\rm age}$) are listed for the rotation-powered model with microphysical parameters $\epsilon_B=0.01$, $\gamma_b=10^5$, $q_1=1.5$ and $q_2=2.5$. The results are tabulated for both USSN ($E_{\rm sn}=10^{50}\,{\rm erg}$, $M_{\rm ej}=0.1\,M_{\odot}$) and CCSN ($E_{\rm sn}=10^{51}\,{\rm erg}$, $M_{\rm ej}=3.0\,M_{\odot}$) progenitors for the respective FRB PRSs.
}
\label{Table1}
\bgroup
\def\arraystretch{1.5}
\begin{tabular}{|c|c|c|c|c|c|}
\hlinewd{0.15pt} \hline
\centering
\textbf{Source} & \textbf{Progenitor} & $(\bm{E_{\rm sn}}$, $\bm{M_{\rm ej}}$) &
$\bm{t_{\rm age}}$ & $\bm{B_{\rm dip}}$ & $\bm{P_{\rm i}}$ \\ \hline 
\hlinewd{0.15pt}
FRB 20121102A & USSN & ($10^{50}\,{\rm erg}$, $0.1\,M_{\odot}$) & $20\,{\rm yr}$ & $3\times10^{12}\,{\rm G}$ & $3.0\,{\rm ms}$ \\ \cline{2-6}
& CCSN & ($10^{51}\,{\rm erg}$, $3.0\,M_{\odot}$) & $32\,{\rm yr}$ & $3\times10^{12}\,{\rm G}$ & $3.0\,{\rm ms}$ \\ \cline{1-6}
FRB 20190520B & USSN & ($10^{50}\,{\rm erg}$, $0.1\,M_{\odot}$) & $20\,{\rm yr}$ & $5\times10^{12}\,{\rm G}$ & $1.5\,{\rm ms}$ \\ \cline{2-6}
& CCSN & ($10^{51}\,{\rm erg}$, $3.0\,M_{\odot}$) & $25\,{\rm yr}$ & $5\times10^{12}\,{\rm G}$ & $1.5\,{\rm ms}$ \\ \cline{1-6}
FRB 20201124A & USSN & ($10^{50}\,{\rm erg}$, $0.1\,M_{\odot}$) & $8\,{\rm yr}$ & $5.5\times10^{13}\,{\rm G}$ & $10\,{\rm ms}$ \\ \cline{2-6}
& CCSN & ($10^{51}\,{\rm erg}$, $3.0\,M_{\odot}$) & $10\,{\rm yr}$ & $5.5\times10^{13}\,{\rm G}$ & $10\,{\rm ms}$ \\ \cline{1-6}
\hlinewd{0.5pt}
\end{tabular}
\egroup
\end{center}
\end{table*}

For a young NS, the magnetised nebula and SN ejecta can be dense enough to provide a large DM inconsistent with observations. DM contribution from the near-source medium should decrease with NS age as the MWN and SN ejecta expand over time. In contrast to the nearly unchangeable DM of FRB 20121102A~\citep{Hessels2019,Oostrum2020,Li2021}, the DM of FRB 20190520B decreases with the rate of $\sim 0.1\,{\rm pc\,cm^{-3}}$ per day~\citep{Niu2022}. We use detailed calculations of ionization and radiation transport for young rapidly rotating NS to predict contributions from the MWN and SN ejecta to FRB DM (see~\citealt{KM2017,MM2018}). 

Figure~\ref{fig:DMevol_Mej_Esn} shows the time evolution of the inferred FRB host galaxy and near source DM contribution, ${\rm DM}_{\rm host}+{\rm DM}_{\rm ns}$, for FRB 20121102A (20190520B) [20201124A] in the left (center) [right] panel. To analyze the DM contributions from the MWN and expanding SN ejecta, for a given progenitor model i.e. USSN/CCSN, we fix the NS parameters to their best-fit values as in Figure~\ref{fig:LC_Mej_Esn}. The host galaxy DM contribution (including its relative uncertainty) is shown in each panel with the shaded yellow region; the shaded magenta region for 20190520B denotes the possibly overestimated ${\rm DM}_{\rm host}$ due to significant contributions from the foreground galaxies and clusters~(see e.g,~\citealt{Lee2023}). As expected, the DM contribution from the SN ejecta and magnetized nebula is substantially higher for CCSN progenitor models as they have a larger $M_{\rm ej}$. The contribution from the magnetised nebula to ${\rm DM}_{\rm ns}$ is almost negligible compared to that from the denser SN ejecta. 

A lower limit on source age $t_{\rm age}$ follows from the requirement that SN ejecta should not be significantly dense to overproduce DM relative to ${\rm DM}_{\rm obs}$. We find the minimum NS age allowed by the near-source DM constraint to be $t_{\rm age,min} \sim 1-3\,{\rm yr}$. The dashed vertical lines and the associated grey-shaded regions in Figure~\ref{fig:DMevol_Mej_Esn} correspond to the condition that the SN ejecta and magnetized nebula should allow for the propagation of radio waves outside the source environment, and therefore be transparent to free-free absorption ($\tau_{\rm ff} \lesssim 1$) and synchrotron self-absorption ($\tau_{\rm sa} \lesssim 1$), respectively. This condition imposes a stronger constraint on minimum NS age with $t_{\rm age,min} \sim 10\,{\rm yr}$ for FRB 20121102A, $\sim 8\,{\rm yr}$ for FRB 20190520B and $\sim 6.5\,{\rm yr}$ for FRB 20201124A. Our inferred best-fit $t_{\rm age}$ from Figure~\ref{fig:SED_Mej_tage} exceeds this minimum NS age required by the absorption criterion.

\subsection{Rotation powered vs magnetic powered models}

\begin{figure*}
\includegraphics[width=0.32\textwidth]{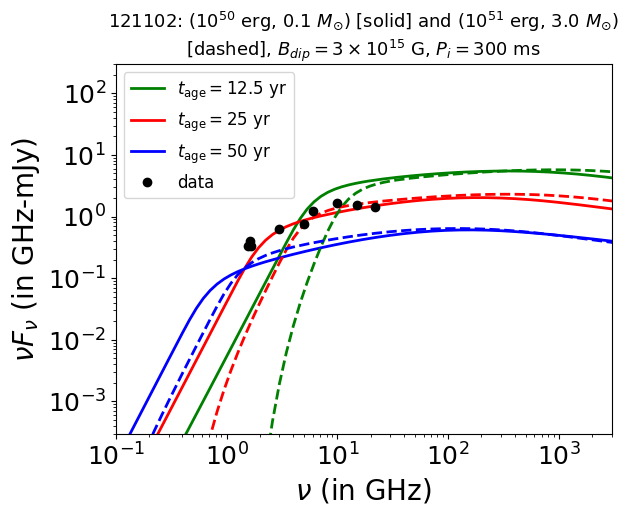}
\includegraphics[width=0.32\textwidth]{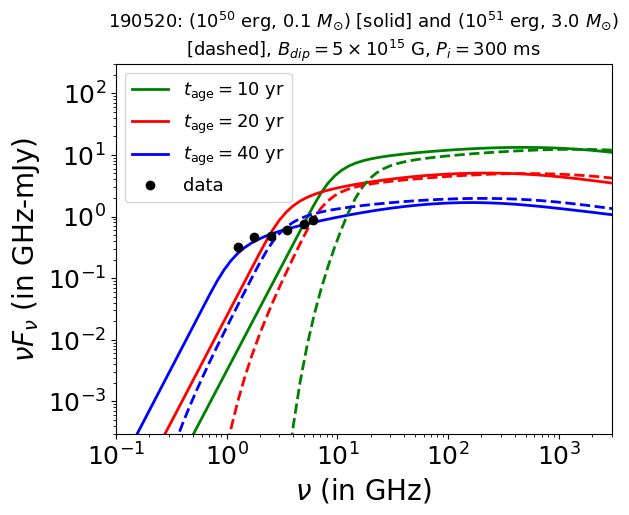}
\includegraphics[width=0.32\textwidth]{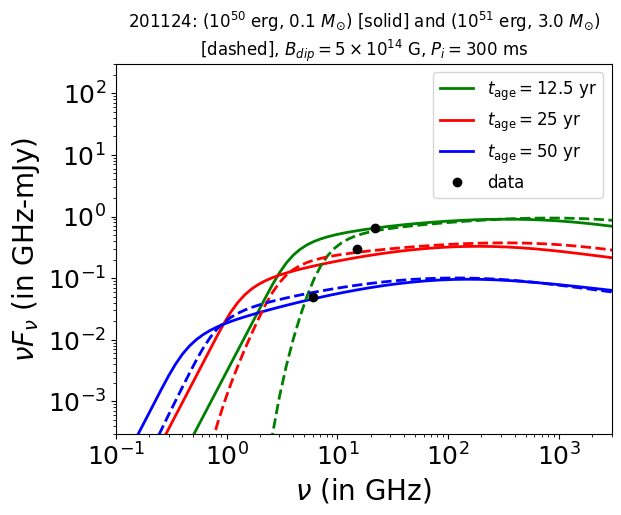}
\caption{Effect of NS age ($t_{\rm age}$) on the radio SED of persistent emission from FRB 20121102A (20190520B) [20201124A] source is shown in the left (center) [right] panel. The fluxes for synchrotron emission are shown for magnetar-flare-powered energy injection model with $\epsilon_B=0.1$, $\gamma_b=10^3$, $q_1=q_2=2.0$. For each source, we fix $P_i=300\,{\rm ms}$ and $B_{\rm dip} \sim (0.01-0.1)\,B_{\rm int} \approx 5\times10^{14}-5\times10^{15}\,{\rm G}$ to vary the NS age within a range that includes the best-fit $t_{\rm age}$. In each panel, the SED data from radio observations is shown with filled circles and the results for USSN/CCSN progenitors are shown using solid/dashed curves. 
} 
\label{fig:Rot_vs_Mag_comp}
\end{figure*}

\begin{figure*}
\includegraphics[width=0.32\textwidth]{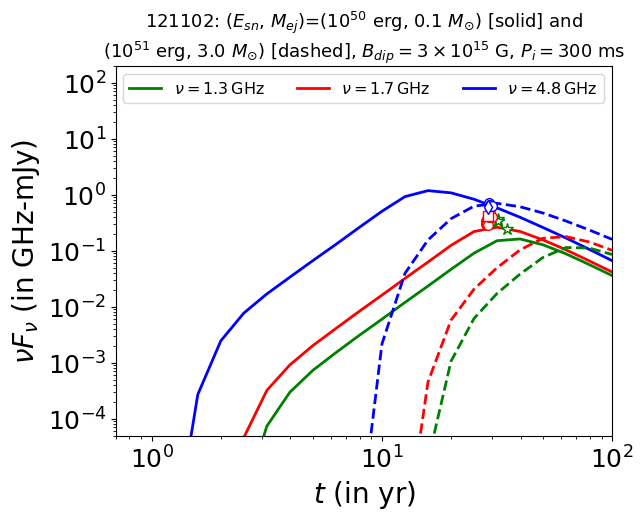}
\includegraphics[width=0.32\textwidth]{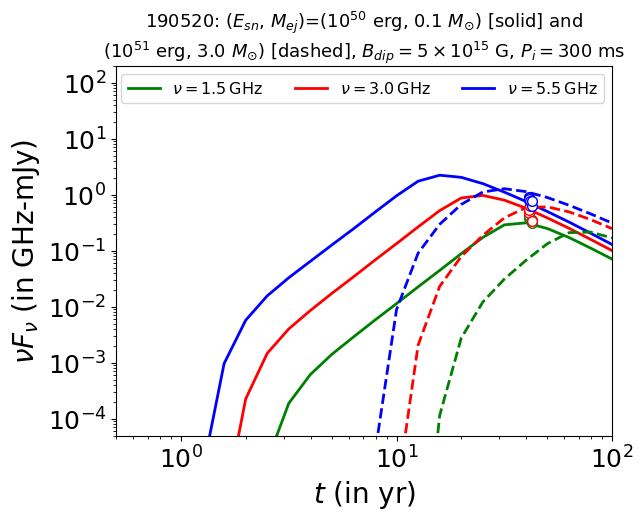}
\includegraphics[width=0.32\textwidth]{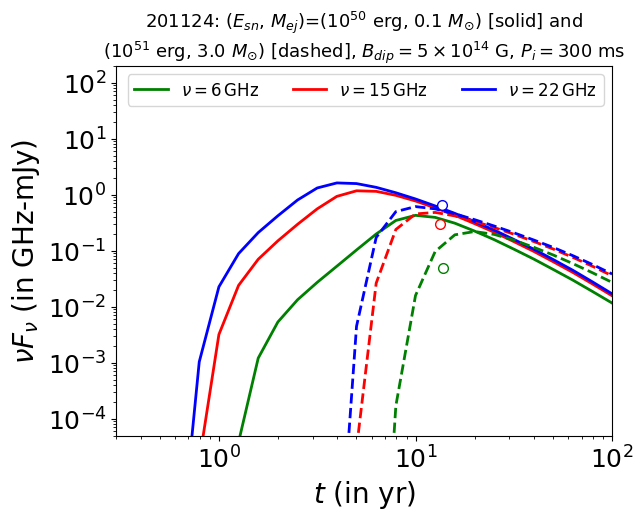}
\caption{Radio light curves for persistent emission associated with FRB 20121102A (20190520B) [20201124A] are shown in the left (center) [right] panel. Corresponding data at various frequencies is shown using unfilled circles for each source. The solid and dashed curves here denote the results for USSN and CCSN progenitors, respectively. We fix the NS parameters ($B_{\rm dip}$, $P_i$, $t_{\rm age}$) to their best-fit values as obtained from Figure~\ref{fig:Rot_vs_Mag_comp} for the magnetar-flare-powered model.} 
\label{fig:LC_Mej_Esn_mag}
\end{figure*}

\begin{table*}
\begin{center}
\caption{Same as Table~\ref{Table1}, but for magnetar-flare-powered model with microphysical parameters $\epsilon_B=0.1$, $\gamma_b=10^3$, $q_1=q_2=2.0$. We set $P_i=300\,{\rm ms}$ for all cases listed here. The corresponding radio SED results for all three sources considered are shown in Figure~\ref{fig:Rot_vs_Mag_comp}.
}
\label{Table2}
\bgroup
\def\arraystretch{1.5}
\begin{tabular}{|c|c|c|c|c|}
\hlinewd{0.15pt} \hline
\centering
\textbf{Source} & \textbf{Progenitor} & $(\bm{E_{\rm sn}}$, $\bm{M_{\rm ej}}$) &
$\bm{t_{\rm age}}$ & $\bm{B_{\rm dip}}$ \\ \hline 
\hlinewd{0.15pt}
FRB 20121102A & USSN & ($10^{50}\,{\rm erg}$, $0.1\,M_{\odot}$) & $25\,{\rm yr}$ & $3\times10^{15}\,{\rm G}$ \\ \cline{2-5}
& CCSN & ($10^{51}\,{\rm erg}$, $3.0\,M_{\odot}$) & $25\,{\rm yr}$ & $3\times10^{15}\,{\rm G}$ \\ \cline{1-5}
FRB 20190520B & USSN & ($10^{50}\,{\rm erg}$, $0.1\,M_{\odot}$) & $40\,{\rm yr}$ & $5\times10^{15}\,{\rm G}$ \\ \cline{2-5}
& CCSN & ($10^{51}\,{\rm erg}$, $3.0\,M_{\odot}$) & $40\,{\rm yr}$ & $5\times10^{15}\,{\rm G}$ \\ \cline{1-5}
FRB 20201124A & USSN & ($10^{50}\,{\rm erg}$, $0.1\,M_{\odot}$) & $12.5\,{\rm yr}$ & $5\times10^{14}\,{\rm G}$ \\ \cline{2-5}
& CCSN & ($10^{51}\,{\rm erg}$, $3.0\,M_{\odot}$) & $12.5\,{\rm yr}$ & $5\times10^{14}\,{\rm G}$ \\ \cline{1-5}
\hlinewd{0.5pt}
\end{tabular}
\egroup
\end{center}
\end{table*}

Energy injected by the rotating magnetar drives the MWN and SN ejecta to evolve together, and the observed quasi-steady emission is powered by the synchrotron radiation generated from the magnetized nebula. For young rapidly rotating NS, the rotational energy is the primary energy reservoir that powers the magnetised nebula and it has been studied in detail for Galactic PWNe~\citep{TT2010,TT2013}. For a typical NS with mass $M_{\rm ns}=1.4\,M_{\odot}$ and radius $R_{\rm ns}=12\,{\rm km}$, the rotational energy is given by $\mathcal{E}_{\rm rot,i} \approx (1.9\times10^{52}\, {\rm erg})\, P_{i,-3}^{-2}$. However, for a magnetar of age $t_{\rm age} \gg t_{\rm sd} \approx (0.12\, {\rm yr})\, B_{\rm dip,13}^{-2} P_{\rm i,-3}^2$, the central NS may spin down considerably over time such that its interior magnetic energy, $\mathcal{E}_{\rm B,int} \approx B_{\rm int}^2 R_{\rm ns}^3/6 = (3\times10^{49}\, {\rm erg})\, B_{\rm int,16}^2$, is comparable or even larger than $\mathcal{E}_{\rm rot,i}$. 

In our analysis, we incorporate models for rotational energy injection proposed by~\citet{Murase2016} and energy injection due to magnetar flares based on~\citet{MM2018}, to compute the total energy injected into the magnetar wind nebula (see equation~\ref{Einj_max}). However, the microphysical parameters $\epsilon_B$, $\gamma_b$, $q_1$ and $q_2$ that determine the actual energy injection rate into the radiating electrons and positrons are given by equation~(\ref{n_e_inj}). In the previous sections, we considered a rotation-powered model with $\epsilon_B=0.01$, $\gamma_b=10^5$, $q_1=1.5$ and $q_2=2.5$. The corresponding best-fit NS parameters ($t_{\rm age}$, $B_{\rm dip}$, $P_i$) are listed in Table~\ref{Table1} for USSN and CCSN progenitor models, based on the radio SEDs obtained for all three sources in Figures~\ref{fig:SED_Mej_tage} and~\ref{fig:SED_Mej_Bdip_Pi}.  

We now consider the scenario where the energy injection from magnetar flares exceeds that from NS rotation ($\mathcal{E}_{\rm B,int} \gtrsim \mathcal{E}_{\rm rot,i}$), with $P_i=300\,{\rm ms}$, $B_{\rm int}=5\times10^{16}\,{\rm G}$ and $B_{\rm dip} \sim 5\times10^{14}-5\times10^{15}\,{\rm G}$. We study the effect of varying $t_{\rm age}$ on the radio SED of synchrotron emission for the magnetar-flare-powered model with $\epsilon_B=0.1$, $\gamma_b=10^3$ and $q_1=q_2=2.0$. The results for FRB 20121102A/20190520B/20201124A are shown in the left/center/right panel of Figure~\ref{fig:Rot_vs_Mag_comp}, with solid/dashed curves denoting USSN/CCSN progenitors and filled circles corresponding to the radio data for each source. As for the rotation-powered model in Figure~\ref{fig:SED_Mej_tage}, the late-time ($t_{\rm age} \gg t_{\rm sd}$) and low-energy ($\nu \lesssim 1\,{\rm GHz}$) synchrotron flux is suppressed due to significant adiabatic losses and SSA, respectively. However, the high-energy radio flux typically attains smaller values compared to the rotation-powered model to then gradually fall off. This is expected as the energy injection into particles occurs at smaller $\gamma_b \sim 5\times10^3$ for the magnetar-flare-powered model. We find that the USSN progenitor model better explains the low-energy ($\nu \lesssim 1\,{\rm GHz}$) radio SED for both FRBs 20121102A and 20190520B, corresponding to best-fit $t_{\rm age} \approx 25\,{\rm yr}$ and $\approx 40\,{\rm yr}$, respectively. On the contrary, the radio SED for FRB 20201124A is only explained by the CCSN model with a best-fit $t_{\rm age} \approx 12.5\,{\rm yr}$ while accounting for the large suppression in radio flux at smaller energies.

For the magnetar-flare-powered model, we show the radio light curves for the persistent emission associated with each FRB source in Figure~\ref{fig:LC_Mej_Esn_mag}. We use the best-fit $B_{\rm dip}$ and $P_i$ values obtained from radio SED fits in Figure~\ref{fig:Rot_vs_Mag_comp} for each FRB source, with the corresponding NS age set to $t_{\rm age}=25\,(40)\,[12.5]\,{\rm yr}$. In contrast to the results obtained in Figure~\ref{fig:LC_Mej_Esn} for the rotation-powered model, we find that the flux predictions from the USSN model are favored by the radio data, especially at smaller frequencies, for both FRBs 20121102A and 20190520B. In case of FRB 20201124A, the radio observations at smaller frequency $\nu \approx 6\,{\rm GHz}$ favor the CCSN model over the USSN progenitor model. These results are consistent with those obtained from the radio SED fits for the magnetar-flare-powered model ($\epsilon_B=0.1$, $\gamma_b=10^3$, $q_1=q_2=2.0$) in Figure~\ref{fig:Rot_vs_Mag_comp}. The best-fit NS parameters ($t_{\rm age}$, $B_{\rm dip}$) for the magnetar-flare-powered scenario with $P_i \approx 300\,{\rm ms}$ are listed in Table~\ref{Table2}, for both USSN and CCSN models, based on the radio SED and light curve results shown in Figure~\ref{fig:Rot_vs_Mag_comp} and~\ref{fig:LC_Mej_Esn_mag}, respectively.

\section{Discussion}
\label{Sec6}
Magnetars and rapidly rotating pulsars have been studied extensively as central engines of SLSNe and FRBs, as they can naturally explain the quasi-steady synchrotron emission from their nascent pulsar/magnetar wind nebulae. These central engines can provide a unified picture of SLSNe, stripped-envelope SNe and FRBs as well as long GRBs. Non-thermal nebular emission has been observed for many Galactic PWNe, which suggests that a significant fraction of the wind magnetic energy can be utilized for accelerating particles. \citet{Murase2016} proposed quasi-steady synchrotron emission as a potential probe of FRB progenitors and also explored their possible connection to pulsar/magnetar-driven SNe including SLSNe. Till date, five repeating FRBs namely 20121102A, 20190520B, 20201124A, 20240114A and 20190417A have been localized and associated with compact persistent radio sources~\citep{Chatterjee2017,Niu2022,Bruni2023,Bruni2024,Moroianu2025}.

In this study, we consider the scenario whereby quasi-steady emission from such FRB sources is powered by young rapidly rotating magnetars that are embedded in a composite of magnetized wind nebula and SN ejecta. We have presented a detailed theoretical model here to numerically compute the radio SEDs for nebular emission by solving the time-dependent kinetic equations for photons and electrons and positrons in the MWN, including inverse Compton and synchrotron emission processes to account for electromagnetic cascades. To test SNe progenitor models, we considered two cases: (a) an ultra-stripped SNe with $M_{\rm ej}=0.1\,M_{\odot}$ and $E_{\rm sn}=10^{50}\,{\rm erg}$, and (b) a conventional core-collapse SNe with $M_{\rm ej}=3.0\,M_{\odot}$ and $E_{\rm sn}=10^{51}\,{\rm erg}$.

While rotational energy from a young NS is likely to be the primary reservoir that powers magnetized nebula, energy released from the dissipation of interior magnetic fields (magnetar flares) can be significant at later times $t_{\rm age} \gg t_{\rm sd}$. We estimated the total energy injected into the MWN based on the NS rotational energy injection model proposed by \citet{Murase2016}. For the magnetar flare energy injection model, we adopted parameters suggested by \citet{Beloborodov2017}. In addition to the SNe progenitor and energy injection models, synchrotron flux from the magnetized nebula is determined by NS parameters such as $B_{\rm dip} \sim 10^{12}-10^{15}\,{\rm G}$, $P_i \sim 1-30\,{\rm ms}$ and $t_{\rm age} \gtrsim t_{\rm obs}$.

We first estimated the allowed combinations of these parameters using physical constraints from: (a) energy injected into the magnetized nebula ($\varepsilon_{\rm nb,min}<\varepsilon_{\rm inj,max}$), (b) near-source DM contribution from the SN ejecta and MWN (${\rm DM}_{\rm ej+nb} < {\rm DM}_{\rm host}$), (c) non-attenuation of radio signal due to free-free absorption and synchrotron self-absorption ($\tau_{\rm ff},\, \tau_{\rm sa} \lesssim 1$), and (d) size of the magnetized nebula ($R_{\rm nb} \lesssim 10\,{\rm pc}$). The corresponding results for $B_{\rm dip}$, $P_i$ and $t_{\rm age}$ are shown in Figure~\ref{FRB_constraint_plots}. We showed that $\varepsilon_{\rm nb,min}<\varepsilon_{\rm inj,max}$ and ${\rm DM}_{\rm ej+nb} < {\rm DM}_{\rm host}$ are the most constraining for the range of parameters and models considered. 
For $t_{\rm age} \gtrsim 10\,{\rm yr}$, the allowed ($P_i$, $B_{\rm dip}$) parameter space generally falls within ($1-10\,{\rm ms}$, $10^{12-14}\,{\rm G}$) for both USSN and CCSN progenitors across the three sources (FRBs 20121102A, 20190520B, 20201124A), with specific ranges detailed in the figure. For both progenitor types and $t_{\rm age} \gtrsim 10\,{\rm yr}$, we found that the parameter space satisfying $L_{\rm sd} > 10^{42}\,{\rm erg/s}$ overlaps with the allowed ($P_i$, $B_{\rm dip}$) parameter space, which indicates that the NS rotational energy is likely to be the primary energy reservoir for these FRB sources detected with persistent radio counterparts. As the flux decline rate is expected to be proportional to the NS spindown timescale, $t_{\rm age}$ can also be inferred from detailed follow-up observations of the PRS.

The localisation of FRBs 20121102A and 20190520B in dwarf host galaxies with high specific star-formation rates~\citep{Chatterjee2017,Tendulkar2017} suggested a connection between FRBs and SLSN progenitors \citep{Perley2016}, as also previously pointed out by~\citet{Murase2016}. Modeling the quasi-steady nebular emission is therefore important to constrain the FRB source properties, as the radio emission can be significantly absorbed in the MWN and SN ejecta at early times ($t_{\rm age} \lesssim {\rm few\, yrs}$). SSA in the magnetized nebula and free-free absorption in the SN ejecta are the primary processes that can suppress the low-energy synchrotron flux, particularly for SNe progenitors with large ejecta mass and/or explosion energy.

Using our numerical code (based on \citealt{Murase2015,Murase2021}) that solves the kinetic equations including electromagnetic cascades, we computed radio SEDs and light curves. We first considered the rotation-powered model with microscopic parameters: $\epsilon_B=1-\epsilon_e=0.01$, $\gamma_b=10^5$, $q_1=1.5$ and $q_2=2.5$. We found that the observed radio flux for both FRBs 20121102A and 20190520B are consistent with an NS of $t_{\rm age} \approx 20\,{\rm yr}$ in a USSN progenitor. This progenitor provides a better match to the larger radio flux at $\nu \approx 1\,{\rm GHz}$ and the harder energy spectra observed for FRBs 20121102A and 20190520B. However, the observed persistent emission for FRB 20201124A, particularly the strong suppression at lower frequencies ($\nu \lesssim 5\,{\rm GHz}$ due to SSA), can only be explained with the CCSN model for a young NS aged $t_{\rm age} \approx 10\,{\rm yr}$, suggesting a larger $M_{\rm ej}$ (see Figure~\ref{fig:SED_Mej_tage}). From our radio SED estimates, we inferred NS parameters: for FRBs 20121102A and 20190520B, $B_{\rm dip} = (3-5)\times10^{12}\,{\rm G}$ and $P_i = 1.5-3\,{\rm ms}$; for FRB 20201124A, $B_{\rm dip} = 5.5\times10^{13}\,{\rm G}$ and $P_i = 10\,{\rm ms}$ (as shown in Figure~\ref{fig:SED_Mej_Bdip_Pi}). An increase in $B_{\rm dip}$ does not necessarily lead to brighter radio emission due to shorter spindown timescales ($t_{\rm sd} \propto B_{\rm dip}^{-2}P_i^2$) causing greater adiabatic energy loss.

We also tested the alternate scenario where energy injection is dominated by magnetar flares using $\epsilon_B=1-\epsilon_e=0.1$, $\gamma_b=10^3$, $q_1=q_2=2$, $P_i=300\,{\rm ms}$, $B_{\rm int}=5\times10^{16}\,{\rm G}$ and $B_{\rm dip}=5\times10^{14-15}\,{\rm G}$. From radio SEDs, we then inferred typical $t_{\rm age}$ such that the MWN can power the persistent radio emission. For FRBs 20121102A and 20190520B, the USSN progenitor model best explains the low-energy radio SED for $t_{\rm age}\approx 25\,{\rm yr}$ and $\approx 40\,{\rm yr}$, respectively. The large suppression in FRB 20201124A's observed radio flux again implied a CCSN progenitor with larger ejecta mass, likely with $t_{\rm age}\approx 12.5\,{\rm yr}$ (see Figure~\ref{fig:Rot_vs_Mag_comp}).

The observed flux density of a PRS can exhibit variability across long timescales that are intrinsic to the source. In a recent study,~\citet{Rhodes2023} reported a $\sim 30\%$ reduction in the flux density of FRB 20121102A's PRS at 1.3 GHz over three years (from 2019 to 2022) with the MeerKAT telescope. Similarly, \citet{Zhang2023} reported a $\sim 20\%$ reduction in the radio flux of FRB 20190520B's PRS from their 3 GHz Very Large Array (VLA) observations in 2020 and 2021. Such long-term variation can be instrumental in constraining the properties of the near-source region. Based on our predicted radio light curves (Figure~\ref{fig:LC_Mej_Esn}), the USSN and CCSN models are hard to distinguish for FRBs 20121102A and 20190520B based on current data. However, recent observations for FRB 20201124A's PRS~\citep{Bruni2023} favor the CCSN model, consistent with our SED results.

While the PRS arises from relativistic plasma, the near-source DM contribution comes primarily from cold plasma (dominated by SN ejecta in our models). The inferred ${\rm DM}_{\rm host}+{\rm DM}_{\rm ns}$ for these FRBs indicates substantial near-source contribution. Our models indicated a minimum NS age $t_{\rm age,min} \sim 1-3\,{\rm yr}$ for these sources such that the SN ejecta does not overproduce the observed DM (see Figure~\ref{fig:DMevol_Mej_Esn}). However, non-attenuation of the radio signal due to free-free absorption in the ejecta and synchrotron self-absorption in the nebula provides a stronger constraint, yielding $t_{\rm age,min} \approx 10\,(8)\,[6.5]\,{\rm yr}$ for FRB 20121102A (20190520B) [20201124A]. The inferred near-source DM is subject to uncertainty from the pair multiplicity in our models \citep{Murase2015}.

Our optimal parameter values ($B_{\rm dip}$, $P_i$, $t_{\rm age}$, $M_{\rm ej}$, $E_{\rm sn}$) are consistent with previous studies~(e.g.,~\citealt{Kashiyama2016,YD2019,ZW2021}). Some works proposed models without SN ejecta \citep{YD2019} or used simplified one-zone models with constant energy injection \citep{CT2024}, which face challenges explaining DM evolution or absorption. Others modelled time-dependent injection, often focusing on magnetar flares and assuming constant nebula/ejecta velocity~\citep{MM2018, ZW2021}. Our analysis improves upon these studies by first deriving constraints on NS and ejecta parameters from observations (DM, energy, PRS size, signal attenuation), then self-consistently computing the synchrotron flux for both NS rotational and magnetic energy injection scenarios, using nebula/ejecta velocities driven by the time-dependent energy injection rate into the MWN.

We note that the present framework focuses primarily on radiative evolution of the magnetar wind nebula, with the dynamical and acceleration components treated in a simplified, phenomenological manner. A fully self-consistent model would require simultaneous treatment of the wind–ejecta interaction and particle acceleration for arbitrary magnetization, which is beyond the scope of this work. Our parameterization in terms of $\epsilon_e$, $\epsilon_B$ and $\gamma_{\rm e,min}$ effectively assumes efficient conversion of upstream Poynting flux to downstream particle energy. The inferred magnetar parameters should therefore be interpreted as conditional upon these assumptions. While quantitative constraints may vary for different dissipation efficiencies or acceleration prescriptions, the qualitative conclusion that young magnetars embedded in synchrotron-emitting nebulae can account for the observed luminosities and spectra of PRSs remains robust.

If repeating FRBs arise from host galaxies that preferentially harbor SLSNe, their rates would be low ($\sim 0.01\%$ of CCSNe, \citealt{Quimby2013}). Conversely, if these were associated with USSN or AICs, rates might be higher ($\sim 0.1-1\%$ of CCSNe, \citealt{Ruiter2009,Tauris2013}). Constraining the number density of repeating FRBs will help distinguish these possibilities. Future searches for radio counterparts of pulsar-driven SN candidates, including SLSNe, especially those powered by rapid NS rotation ($P_i \sim 1-30\,{\rm ms}$), will give crucial information on their relationship with FRBs. Such non-thermal nebular emission should be distinguishable from Galactic thermal sources.

Recently,~\citet{Xing2024} reported a GeV gamma-ray flare associated with the hyper-active repeating FRB 20240114A. While our MWN model allows for such emission in the forward shock region~\citep{Murase2016}, reproducing the observed gamma-ray flux requires more efficient magnetic energy dissipation than assumed here. 
Furthermore, hard X-ray emission can also serve as another promising signal for parameters accounting for both SLSNe and FRBs~\citep{MP2014,Kashiyama2016}. While some X-rays might be reprocessed by SN ejecta into thermal optical/UV transient lasting over days, the optical depth of pairs in the nebula could suppress thermalization. If nebular X-rays are energetic enough to reionize the ejecta, non-thermal X-rays could escape unattenuated with a peak luminosity and timescale similar to optical radiation.

\section{Summary}
\label{Sec7}
This study investigated the origin of PRSs associated with three localized repeating FRBs: 20121102A, 20190520B, and 20201124A. Recently, two more confirmed FRB-PRS associations have been reported for FRBs 20240114A and 20190417A.
Motivated by the potential for young and rapidly rotating magnetars powering wind nebulae (MWN) to explain various energetic transients such as superluminous supernovae (SLSNe), we explored the scenario where the observed persistent radio emission arises due to quasi-steady synchrotron radiation from an MWN expanding within the SN ejecta that birthed the magnetar.

We employed a detailed theoretical model, numerically solving kinetic equations for particles and photons within the MWN to predict radio SEDs and light curves, considering emission/absorption processes and electromagnetic cascades. Both NS spindown and magnetar flares were treated as possible energy sources. Our model is tested against two distinct SN progenitor environments – an ultra-stripped SN with low ejecta mass and a conventional core-collapse SN with high ejecta mass – while constraining key NS parameters ($P_i, B_{\rm dip}, t_{\rm age}$) using physical limits derived from energy, dispersion measure, radio attenuation, and nebula size.

Our results indicated that the NS rotational energy is generally sufficient to power observed PRS luminosities. Comparing model predictions with observations, we found the properties of FRBs 20121102A and 20190520B (higher flux at lower frequencies, harder spectra) are better matched by a magnetar aged $\sim 20$ yr within a USSN environment. Conversely, the distinct low-frequency turnover in FRB 20201124A's spectrum points towards significant self-absorption, favoring a younger magnetar ($\sim 10-12.5$ yr) residing in the denser ejecta of a CCSN. These findings are consistent with minimum age estimates of $\sim 7-10$ yr derived from radio transparency constraints, lending support to the MWN model as a viable explanation for FRB persistent radio counterparts and potentially linking some FRBs to SN progenitor channels.

\section*{Acknowledgements}
We thank Ke Fang, Pawan Kumar and Navin Sridhar for useful discussions and comments. We also thank the anonymous reviewer for their helpful comments and suggestions that improved the quality of this manuscript.
M.B. acknowledges support from the Eberly Research Fellowship at the Pennsylvania State University and the Simons Collaboration on Extreme Electrodynamics of Compact Sources (SCEECS) Postdoctoral Fellowship at the Wisconsin IceCube Particle Astrophysics Center (WIPAC), University of Wisconsin-Madison. 
The work of K.M. is supported by the NSF Grant No.~AST-2108466, No.~AST-2108467, and AST-2308021, and KAKENHI No.~20H05852.
The work of K.K. is supported by JSPS KAKENHI Grant Nos.~JP24K00668,~JP23H04899,~JP22H00130.

\section*{Data availability}
The data underlying this article will be shared on reasonable request to the corresponding author.

\bibliography{main}

\label{lastpage}

\end{document}